\documentstyle[twoside,psfig,11pt]{article}
\makeindex

\input{psfig.sty}

\newcommand{\epsFig}[2]{
\begin{figure}
\label{#1}
\centerline{\leavevmode
\psfig{figure=./#1.eps}}
\caption{#2}
\end{figure}
}

\newcommand{\qed}{\vspace{.1em}\noindent\fbox{\rule{
0em}{.1em}\rule{.1em}{0em}}\vspace{1em}}

\newcommand{\definitionBullet}{$\bullet$}

\newenvironment{proof}{

\noindent{\bf Proof:}\ }{
\hfill \qed

}

\newtheorem{theorem}{Theorem}[section]
\newtheorem{lemma}[theorem]{Lemma}
\newtheorem{corollary}[theorem]{Corollary}

\newcommand{\lecture}[1]{
\cleardoublepage
\markright{}
\section{#1}
\markright{\thesection---#1}
}

\newcommand{\topic}[1]{\subsection{#1}}

\newcommand{\subtopic}[1]{\subsubsection{#1}}

\newcommand{\note}[1]{{\marginpar{\footnotesize #1}}}

\newcommand{\define}[1]{{\em #1}\index{#1}\note{#1}}
                
\newcommand{\defnote}[2]{\footnote{#2}\index{#1}\note{#1}}

\newcommand{\restrict}[2]{{\left.#1\right|_{#2}}}

\newcommand{\R}{\rm I\!R}               
\newcommand{\conv}{{\ {\rm conv}}}      
\newcommand{\affine}{{\ {\rm affine}}}  
\newcommand{\minus}{\backslash}         
\newcommand{\link}{/}                   
\newcommand{\GF}{{\rm GF}}              
\newcommand{\divides}{{\,\mid\,}}       
\newcommand{\mod}{\ {\rm mod}\ }        
\newcommand{\Mod}{{\rm mod}\ }          
\newcommand{\mobius}{{\cal M}}          
\newcommand{\K}{{\cal K}}               
\newcommand{\C}{{\cal C}}               
\newcommand{\fix}{{\rm fix}}            
\newcommand{\h}{{\cal H}}               
\newcommand{\A}{{\cal A}}               
\newcommand{\p}{{\cal P}}               
\newcommand{\transpose}{^{\rm T}}       

\begin{document}

\title{
Lecture Notes on \\
Evasiveness of Graph Properties
}
\author{
\normalsize Lectures by \\
{\large \it L\'aszl\'o Lov\'asz} \\
\normalsize Computer Science Department, Princeton University, \\
\normalsize Princeton, NJ 08544, USA, \\
\normalsize and Computer Science Department, E\~otv\~os Lor\'and University, \\
\normalsize Budapest, Hungary H-1088.
\and
\normalsize Notes by \\
{\large \it Neal Young} \\
\normalsize Computer Science Department, Princeton University, \\
\normalsize Princeton, NJ 08544, USA. \\
}
\date{}
\maketitle
\begin{abstract}
These notes cover the first eight lectures of the class {\it Many Models of
Complexity} taught by L\'aszl\'o Lov\'asz at Princeton University in the Fall
of 1990.  The first eight lectures were on evasiveness of graph properties and
related topics; subsequent lectures were on communication complexity and
Kolmogorov complexity and are covered in other sets of notes.

The fundamental question considered in these notes is, given a function, how
many bits of the input an algorithm must check in the worst case before it
knows the value of the function.  The algorithms considered are deterministic,
randomized, and non-deterministic.  The functions considered are primarily
graph properties --- predicates on edge sets of graphs invariant under
relabeling of the edges.  
\end{abstract}
\newpage
\tableofcontents

\lecture{Decision Trees and Evasive Properties}

The goal of this course is to examine various ways of measuring the
complexity of computations.  In this lecture, we discuss the decision
tree complexity of functions.  We begin a characterization of which
functions require that for any deterministic algorithm for computing
the function, there is some input for which the algorithm checks all
the bits of the input.

\topic{Decision Trees}
A \define{decision tree} is a tree representing the logical structure
of certain algorithms on various inputs.  The nodes of the tree
represent branch points of the computation --- places where more than
one outcome are possible based on some predicate of the input --- and
the leaves represent possible outcomes.  Given a particular input, one
starts at the root of the tree, performs the test at that node, and
descends accordingly into one of the subtrees of the root.
Continuing in this way, one reaches a leaf node which represents the
outcome of the computation.

\epsFig{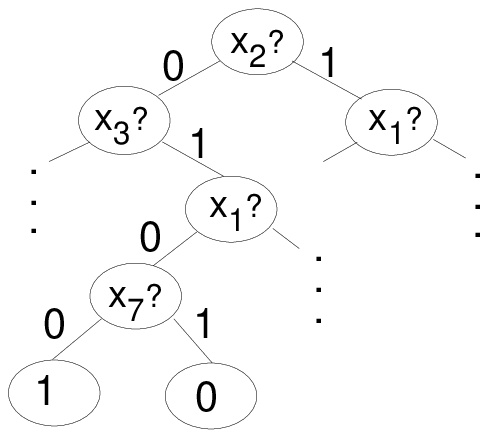}{A Simple Decision Tree}

Given a function $f:\{0,1\}^n \rightarrow \{0,1\}$, a \define{simple
decision tree} for the function is a binary tree whose internal nodes
have labels from $\{1,2,\ldots,n\}$ and whose leaves have labels from
$\{0,1\}$.  If a node has label $i$, then the test performed at that
node is to examine the $i$th bit of the input.  If the result is 0,
one descends into the left subtree, whereas if the result is 1, one
descends into the right subtree.  The label of the leaf so reached is
the value of the function on the input.

While it is clear that any such function $f$ has a simple decision tree, we
will be interested in simple decision trees for $f$ which have minimal
depth\index{$D(f)$} $D(f)$.  $D(f)$ is called the \define{decision tree
complexity of $f$}.

It is clear that $D(f)$ is at most the number of variables of $f$.  A
simple example which achieves this upper bound is the parity function
$f(x_1,\ldots,x_n) = x_1 + x_2 + \cdots + x_n \mod 2$.  For this
function every leaf of any simple decision tree for $f$ has depth $n$,
because if the value of some $x_i$ has not been examined by the time a
leaf is reached for an input $x$, the tree gives the same answer when
$x_i$ is flipped, so the function computed is not parity.

\topic{An Evasive Function}
A function $f$ with $D(f)$ equal to the number of variables is said to
be \define{evasive}.  A less trivial example of an evasive function is
\[f(x_{ij} : i,j \in \{1,\ldots,n\})
 = \bigwedge_i \bigvee_j x_{ij},\] that is $f$ is 1 iff every row of
the matrix with entries $x_{ij}$ has at least one 1.

To show $f$ is evasive, we use an \define{adversary argument}.  We
simulate the computation of some decision tree, except instead of
checking the bits of the input directly, we ask the adversary.  The
adversary, when asked for the value of $x_{ij}$, responds 0 as long
as some other variable in the row remains undetermined, and 1
otherwise.  In this way the adversary maintains that the value of the
function is undetermined until all variables have been checked.  Note
that in the case we show only that {\em some} leaf is of depth $n^2$.

\topic{A Non-Evasive Function}
Next we give a non-trivial example of a non-evasive function.  Given
players $1,2,\ldots,n$, let $x_{ij} : 1 \le i < j \le n$ be 1 if
player $i$ will beat player $j$ if they play each other and 0 if $j$
will beat $i$.  (No draws allowed.  Note that this is not necessarily
a transitive relation.)  The function is 1 iff there is some player
who will beat everyone.

The object is to determine $f$ without playing all possible matches.
To do this, first play a ``knockout tournament'' --- have 1 and 2
play, have the winner play 3, have the winner play 4, etc.\ until
every player but some player $i$ has lost to somebody.  Now play $i$
against everyone he hasn't played.  If $i$ wins all his matches, $f$
is 1, otherwise $f$ is 0.  The number of matches played in the first
stage is $n-1$, and at most $n-2$ are played in the second, so $D(f)
\le 2n-3$.  (How can one redesign the first stage to show $D(f)
\le 2n - \lfloor\log_2 n\rfloor$?)

\topic{Non-Deterministic Complexity}
The basic idea behind these two examples is that for most functions
(what are the two exceptions?) there are proper subsets of the
variables whose values can determine the value of the function
irrespective of the values of the other variables.  The goal in
minimizing decision tree depth is to discover the partial assignments
as quickly as possible, while the goal in showing large decision tree
complexity is to show this is not possible.  Define\index{$D_0(f), D_1(f)$}
\begin{eqnarray*}
D_1(f) & = & \max_{x:f(x)=1}\min
\{k : \exists i_1,\ldots,i_k,\epsilon_1,\ldots,\epsilon_k :
\restrict{f}{x_{i_1} = \epsilon_1, \ldots, x_{i_k} = \epsilon_k} \equiv 1\}, \\
D_0(f) & = & \max_{x:f(x)=0}\min
\{k : \exists i_1,\ldots,i_k,\epsilon_1,\ldots,\epsilon_k :
\restrict{f}{x_{i_1} = \epsilon_1, \ldots, x_{i_k} = \epsilon_k} \equiv 0\}.
\end{eqnarray*}
That is, $D_i(f)$ is the least $k$ so that from every assignment we
can pick $k$ variables such that assigning only these $k$ values
already forces the function to be $i$.  Alternatively, $D_i(f)$
corresponds to the non-deterministic decision tree complexity of
verifying $f(x)=i$, and $\max\{D_0(f),D_1(f)\}$ is the
non-deterministic decision tree complexity of computing $f$.  (A
non-deterministic computation may be considered as an ordinary
computation augmented by the power to to make lucky guesses.)

\topic{$D(f) \le D_0(f) D_1(f)$}
For boolean $x$ let $x^\epsilon$ denote $x$ if $\epsilon=0$ and
$\bar{x}$ if $\epsilon = 1$.  The representation
\[f(x_1,\ldots,x_n) = \bigvee_l \bigwedge_{i \in S_l} x_i^{\epsilon_{il}}\]
of $f$ in terms of the disjunction of a number of elementary
conjunctions of literals\defnote{literal}
{A {\em literal} is a boolean variable or its negation.}
is called a \define{disjunctive normal form} (DNF) of $f$.  

If we can represent $f$ in DNF so that every elementary conjunction
has at most $k$ terms, then $D_1(f) \le k$, because if any partial
assignment of variable forces $f$ to be 1, it must force some
elementary conjunction to be 1.  Conversely, there exists a DNF for
$f$ in which every elementary conjunction has at most $D_1(f)$ terms:
for $\epsilon:f(\epsilon) = 1$ let $S_\epsilon$ be the indices of
the minimum set of (at most $D_1(f)$) variables whose assignment
$x_i=\epsilon_i$ forces $f$ to 1.  Then
\[f(x) = \bigvee_{\epsilon:f(\epsilon)=1}
\bigwedge_{i\in S_\epsilon}x_i^{1-\epsilon_i}.\]

One can similarly correlate $D_0$ and the \define{conjunctive normal
form} CNF of $f$.

Next, we show the surprising relation $D(f)\le D_1(f)D_0(f)$.  Write
$f$ simultaneously in DNF and CNF so that the sizes of the elementary
conjunctions (disjunctions for CNF) do not exceed $D_1(f)$ ($D_0(f)$).
To determine the value of $f$ on an input $x$, we use the following
strategy.  We choose the first variable $x_i$ in the first elementary
conjunction of the DNF, and query its value $\epsilon_i$.  We then
substitute the value $\epsilon_i$ for the variable $x_i$ in the DNF
and the CNF and simplify, obtaining a DNF and CNF for
$f'=\restrict{f}{x_i=\epsilon_i}$.  Since each elementary conjunction
in the new DNF has size at most $D_1(f)$, $D_1(f') \le D_1(f)$.
Similarly, $D_0(f') \le D_0(f)$.

The crucial observation is that {\em each elementary disjunction in
the CNF has a variable (in fact a literal) in common with each
elementary conjunction in the DNF.} (Otherwise the variables in the
elementary disjunction and the elementary conjunction can be
simultaneously set to force the function to 0 and 1.)  Thus by
continuing the above process, by the time we have queried all of the
at most $D_1(f)$ variables in the first elementary conjunction, we
have reduced the size of every elementary disjunction by at least 1.
It follows that we can query at most the variables in the first
$D_0(f)$ elementary conjunctions before we have determined the value
of the function.  Thus $D(f)\le D_0(f)D_1(f)$.

Recalling the earlier remark about non-determinism, $D_1(f)$, and
$D_0(f)$, one might say that the above shows that in this model
${\rm NP}\cap{\rm co-NP}={\rm P}$.

\topic{The Aanderaa-Karp-Rosenberg Conjecture}
We can represent functions on graphs by encoding the adjacency matrix
in the input to the function.  For an undirected graph with $n$ nodes,
we let $x^G_{ij} : 1\le i < j \le n$ represent the presence or absence
of the edge $(i,j)$ by taking the value 1 or 0 respectively.  

In this way we can represent arbitrary functions on graphs.
Generally, however, we will restrict our attention to \define{graph
properties} --- boolean functions whose values are independent of the
labeling of the nodes of the graph.  Technically, 
$f:\{x_{ij} : 1 \le i < j \le n\}\rightarrow\{0,1\}$
is a graph property if for any
$\Pi \in S_n$,\defnote{$S_n$}
{$S_n$ denotes the symmetric group on $n$ elements, also
known as the set of permutations of size $n$} 
and for any $x$,
\[f(\ldots,x_{ij},\ldots) = f(\ldots,x_{\Pi(i)\Pi(j)},\ldots).\]

The Aanderaa-Karp-Rosenberg\index{Aanderaa}\index{Karp}\index{Rosenberg} (AKR)
Conjecture\index{AKR Conjecture} is that any monotone\defnote{monotone}{A
graph property is {\em monotone} if adding edges to the graph preserves the
property.} , non-trivial graph property is evasive.  It is known to be true
for $n$ a prime power, and counter-examples are known if the monotonicity
requirement is dropped.

A generalization of this conjecture follows.  $F$ is \define{weakly
symmetric} if there exists a transitive\defnote{transitive}
{$G$ is {\em transitive} if $\forall i,j \exists g\in G : g(i) = j$.} 
group $G\subseteq S_n$ such that for all $g\in G$,
$f(\ldots,x_i,\ldots)=f(\ldots,x_{g(i)},\ldots)$.  The generalized
conjecture is that any monotone, non-trivial, weakly symmetric
boolean function is evasive.

For example, suppose $f \equiv$ ``graph $G$ has no isolated node''.
First, observe that for general $f$, if $\#\{x\in\{0,1\}^n:f(x)=1\}$
is odd, then $f$ is evasive.  To see this, observe that for any $x_i$
the above property is maintained for either $\restrict{f}{x_i=0}$ or
$\restrict{f}{x_i=1}$, so that the adversary can answer queries so as
to maintain the property as $f$ is restricted.  As long as the number
of unqueried variables is at least 1, the size of the range of the
restricted function is even, so the property ensures that the function
is not constant.

For the above choice of $f$, an inclusion/exclusion argument shows
\begin{eqnarray*}
\#\{G:G{\rm\ has\ no\ isolated\ vertex}\}
& \equiv & \sum_{k=0}^n (-1)^{k-1}{n \choose k}2^{n-k \choose 2} (\Mod 2)\\
& \equiv & (-1)^{n-1}n + (-1)^n (\Mod 2).
\end{eqnarray*}
Thus provided $n$ is even, an odd number of graphs have no isolated
nodes, and $f$ is evasive.

Note for later that we can generalize the above condition.  In
particular, an inductive argument in the same spirit shows:
\begin{lemma}
\[2^{n-D(f)} \divides \#\{x : f(x) = 1\}.\]
\end{lemma}
\lecture{Evasiveness, continued}

\topic{Connectivity is Evasive}
If $f\equiv$ ``$G$ is connected'', then $f$ is evasive.  To see this
have the adversary answer ``no'' unless that answer would imply that
the graph was disconnected, in which case she answers ``yes''.  In
this way the adversary maintains that a spanning tree exists among the
``yes'' and unqueried edges.  If some edge $(i,j)$ has not been
queried, can the answer be known?  If it is known, it must be ``yes'',
and the ``yes'' edges must contain a spanning tree, so a path of
``yes'' edges connects $i$ to $j$.  Of the edges on this path, suppose
the last edge queried is $(u,v)$.  At this point, we have a contradiction,
because the the adversary could have answered ``no'' to the query of
$(u,v)$ while maintaining the possible connectedness of the graph
through the other ``yes'' edges and edge $(i,j)$.

Consideration shows this argument generalizes to any monotone $f$ with
the property that for any $x$ such that $f(x)=1$, and any $x_i=1$, we
can set $x_i=0$, possibly setting some other $x_j=1$, without changing
the value of the function.

\topic{``Tree'' Functions are Evasive}
A general class of simple but evasive functions are \define{tree
functions} --- those which have formulas using $\vee$ and $\wedge$ in
which every variable occurs exactly once.  The adversary has the
following strategy.  When asked for the value of $x_i$, if $x_i$
occurs in a conjunction ($\cdots\wedge x_i\wedge\cdots$) in the
formula, the adversary claims $x_i=1$.  Otherwise $x_i$ occurs in a
disjunction and the adversary responds that $x_i=0$.  The adversary
plugs the answered value into the formula, simplifies it, and
continues.  In this way, the adversary maintains that one variable is
removed from the formula with each question, so the result can not be
known unless every variable has been queried.  (Clearly the same proof
applies if the formula also contains negations.)

\topic{The AKR Conjecture is True for Prime $n$}
\index{AKR Conjecture}
Previously we proved that if the number of $x$ with $f(x)=1$ is odd,
then $f$ is evasive, and noted that this can be generalized to show
that $2^{n-D(f)}$ divides this number.  Here is an alternate extension:
let $|x|$ denote the number of 1's in $x$.  Define\index{$\mu(f)$}
\[\mu(f) = \sum_{f(x)=1}(-1)^{|x|}.\]  
Then we can use the property
$\mu(f)=\mu(\restrict{f}{x_i=0})-\mu(\restrict{f}{x_i=1})$ to show
that if $\mu(f)\neq0$ then $f$ is evasive.  In particular, the
adversary maintains that $\mu$ applied to the restricted function
(i.e.\ $f$ restricted by the partial assignment given by the
adversary's responses so far) is non-zero, so that the restricted
function is non-trivial unless all variables have been queried.  (The
reader may want to check the base case of this argument.)

More generally, define\index{$p_f(t)$} $p_f(t)=\sum_xf(x)t^{|x|}$.  Then for a
constant function $c$ of $k$ variables, $p_c(t)=(1+t)^k$, and so an inductive
argument similar to the above shows
\[(t+1)^{n-D(f)}\divides p_f(t).\]

Next we use the $\mu$ criterion to prove the generalization of the AKR
conjecture for prime $n$.  A counter-example exists with $n=14$ when
$n$ is not required to be prime.
\begin{theorem}\index{AKR conjecture, proof for prime $n$}
\label{weakSymThm}
If $f:\{0,1\}^n\rightarrow\{0,1\}$ is weakly symmetric\index{weakly
symmetric}, $f(\underline{0})\neq f(\underline{1})$, and $n$ is prime, then
$f$ is evasive.  
\end{theorem}
\begin{proof}
We will show $\mu(f) = \sum_xf(x)(-1)^{|x|} \neq 0$.  The first part
of the proof is to use the weak symmetry of $f$ and the primality of
$n$ to show that there is a permutation consisting of a single cycle
leaving $f$ invariant.  The second is to use this fact and the
primality of $n$ to group the inputs yielding $f(x)=1$ except 0 or 1
into equivalence classes of size $n$, thus showing that 
\(\mu(f)\equiv 1(\Mod n)\),
so that $\mu(f) \neq 0$.

Since $f$ is weakly symmetric, there exists a transitive subgroup $\Gamma$ of
$S_n$ leaving $f$ invariant.  Consider the partition of
$\Gamma=U_1\cup\cdots\cup U_n$ where $g\in U_i$ iff $g(1) = i$.  The
transitivity of $\Gamma$ ensures that each $U_i$ is of the same size, so $n$
divides $|\Gamma|$.  Since $n$ is prime and $n\divides|\Gamma|$, Cauchy's
theorem\index{Theorem, Cauchy} implies that $\Gamma$ contains an element
$\gamma$ of order $n$.  Since $n$ is prime, such a permutation necessarily
consists of a single cycle.

Now (assuming WLOG that $f(0)=0$) we partition the inputs $x$ into
classes such that two elements are in the same class iff one is
obtainable from the other by rotation (i.e.\ application of $\gamma$).
Since $\gamma$ leaves $f$ invariant, and $n$ is prime, it follows that
unless every $x_i$ is the same, each of the $n$ possible rotations of
$x$ are distinct.  Thus the values of $x$ such that $f(x)=1$ can be
partitioned into classes of size $n$, except for $x=\underline{1}$.
It follows that the number of such inputs modulo $n$ is 1, so that the
number of distinct non-zero terms in the expression for $\mu$ is 1
modulo $n$, and $\mu$ is not zero.
\end{proof}

[Here is a sketch of how to generalize the theorem for $n=p^a$ a prime
power.  It is no longer necessarily true that $\Gamma$ has a cyclic
element, but now $\Gamma$ has a transitive (sylow) subgroup $\Gamma'$
of order $p^b$, with $p^b$ but not $p^{b+1}$ dividing $|\Gamma|$.

Again we group the terms of $\mu(f)$ so that two $x$'s are in the
same group if mapped by $\Gamma'$ to each other.  We look at the
orbits of $\Gamma'$ acting on $\{0,1\}^n$.  The number of elements in
an orbit divides $\Gamma'=p^b$ and is not equal to 1 unless
$x=\underline{0}$ or $x=\underline{1}$, and all vectors in the same
orbit give the same value of $f$.

Using this grouping we show $\mu(f) \equiv (-1)^n \mod p$, so
$\mu(f)\neq 0$.]

Before we observed that 
\((t+1)^{n-D(f)} \divides p_f(t)=\sum_xf(x)t^{|x|}\).
We define\index{$p_f(t_1,\ldots,t_n)$}
$p_f(t_1,\ldots,t_n)=\sum_xf(x)t_1^{x_1}\cdots t_n^{x_n}$, and generalize this
observation in the next lemma.

\begin{lemma} 

\(p_f \in
\left\langle(t_{i_1}+1)\cdots(t_{i_{n-D(f)}}+1)
 : 1\le i_1<\cdots<i_{n-D(f)}\le n\right\rangle
\)\defnote{ideal}
{$\langle\cdots\rangle$ represents the ideal generated by $\cdots$
(the smallest set of polynomials closed under subtraction and under
multiplication by any polynomial).  An equivalent formulation of this
lemma is
\[p_f = \sum_{1\le i_1<\cdots<i_{n-D(f)}\le n}
P_{i_1,\ldots,i_{n-D(f)}}\times(t_{i_1}+1)\cdots(t_{i_{n-D(f)}}+1),\]
where the $P_{\cdots}$ are integer coefficient
polynomials of the $t_i$.}

\end{lemma}

\begin{proof}
\epsFig{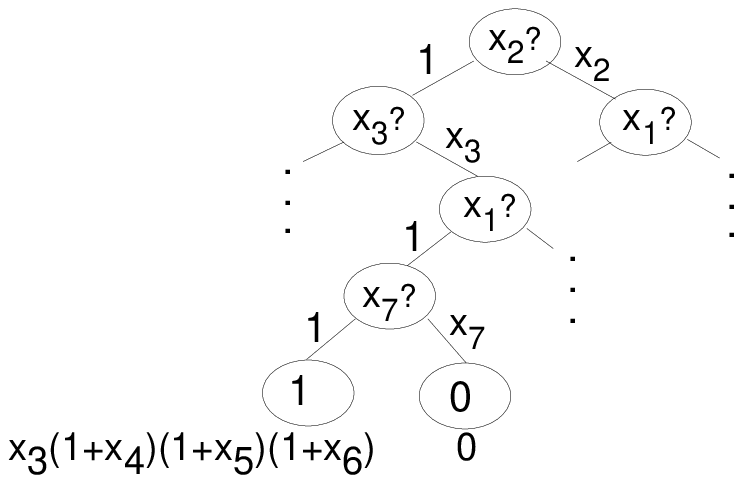}{A term of $P_f$}
First, if $f \equiv 0$, $p_f=0$, and if $f \equiv 1$, 
\(p_f=\sum_x t_1^{x_1}\cdots t_n^{x_n} = (t_1+1)\cdots(t_n+1)\).  
If $f$ is not constant, fix a minimum depth decision tree for $f$ and
use $p_f=p_\restrict{f}{x_1=0}+t_1p_\restrict{f}{x_1=1}$ to expand
$p_f$ into a sum of terms, each term corresponding to a ``yes'' leaf
of the tree.  Each such term is of the form 
\(\left(\Pi_{i\in S_1}t_i\right)\times
\left(\Pi_{i\in \overline{S}} (t_i+1)\right)\),
where $S_1$ is the set of indices of variables queried and found to be
1, and $S$ is the set of variables queried.
\end{proof}

Here is another way to look at this result.  Let $z_i = t_i+1$, and
\begin{eqnarray*}
Q_f(z_1,\ldots,z_n)
& = & p_f(z_1-1,\ldots,z_n-1) \\
& = & \sum_xf(x)(z_1-1)^{x_1}\cdots(z_n-1)^{x_n} \\
& = & \sum_xf(x)\sum_{y\le x}z_1^{y_1}\cdots z_n^{y_n}\times(-1)^{|x-y|} \\
& = & \sum_y\left(\sum_{x\ge y}(-1)^{|x-y|}f(x)\right)
z_1^{y_1}\cdots z_n^{y_n} 
\end{eqnarray*}
(Here the inequality $y\le x$ means $\forall i, y_i \le x_i$.)

By the previous lemma, the terms in $Q_f$ have degree at least
$n-D(f)$ in the $z_i$.  Thus if $|y|< n-D(f)$, then
$\sum_{x\ge y}(-1)^{|x-y|}f(x) = 0$.  The left hand side of this
equality is known as the \define{m\"obius transform} $\mobius f$ of $f$.
Thus we have:
\begin{corollary}
For $|y|\le n-D(f)$, $(\mobius f)(y) = 0$.
\end{corollary}

\lecture{Non-Evasive Monotone Properties Give Contractable Complexes}

In the previous lecture, we showed that every non-trivial 
weakly symmetric function on $p^k$ variables for prime $p$ 
is evasive.

In this lecture we continue our study of evasiveness, introducing some
topological concepts related to simplicial complexes.  We show that the
simplicial complex associated with a non-evasive monotone function is
contractable.  This is the first part of a technique due to Kahn,
Saks, and Sturtevant; our goal is to prove that
all non-trivial, monotone, bipartite graph properties\footnote{ A bipartite
graph property $f(x_{ij} : i\in V,j\in W)$ is a boolean function invariant
under permutations of the edges induced by permutations of $V$ and $W$.  See
``graph property''.} and non-trivial, monotone graph properties of graphs with
a prime power number of nodes are evasive.

\topic{Simplicial Complexes}

A \define{simplicial complex} is a finite collection $\K$ of sets such that
\begin{enumerate}
\item \(\forall X \in \K, Y \subseteq X \Rightarrow Y \in \K\), and
\item \(\K \neq \emptyset\).
\end{enumerate}
\define{$V(\K)$}, the vertices of $\K$, consists of the elements of
the sets in $\K$.  

Corresponding to $\K$ one can construct a geometric
realization\index{$\widehat{\K}$} $\widehat{\K}\subseteq \R^{V(\K)}$.  First,
one defines the mapping $\widehat{\cdot}:V(\K)\rightarrow\R^{V(\K)}$ so that
no vertex is mapped into the affine hull\defnote{$\affine\{S\}, \conv\{S\}$}{
$\affine\{S\}=\left\{\sum_{v\in S}\alpha_v v: \sum \alpha_v=1\right\}$;
$\conv\{S\}=\left\{\sum_{v\in S}\alpha_v v: \sum\alpha_v=1,\alpha_v\ge
0\right\}$.} of any other subset of the vertices (for instance, one maps the
vertices to the unit vectors).  Then one extends $\widehat{\cdot}$ to any set
$X$ of vertices by $\widehat{X}=\conv\{\widehat{v}:v\in X\}$, and to any
collection $\C$ (such as $\K$) of sets by
$\widehat{\C}=\cup_{X\in\C}\widehat{X}$.

Note that for any $X\in\K$, $\widehat{X}$ is a \define{simplex} ---
the convex hull of a set of vectors none of which lies in the affine
hull of any subset of the others.  (Such a set of vectors is said to
be \define{affinely independent}.)

A collection $\K=\{S_1,\ldots,S_m\}$ of simplices in $\R^N$ is said to
form a \define{geometric simplicial complex} if:
\begin{enumerate}
\item  $\forall S_i\in\K$, $T$ a 
face\defnote{face}
{A {\em face} of a simplex $S=\conv\{V\}$ is
a set $S'=\conv\{V'\} : V'\subseteq V$.}
of $S_i \Rightarrow T\in\K$, and
\item $\forall S_i,S_j\in\K,S_i\cap S_j\neq\emptyset\Rightarrow
S_i\cap S_j$ is a face of both $S_i$ and $S_j$.
\end{enumerate}

A \define{polyhedron} is then defined as the union of the sets in any
geometric simplicial complex.  Note that such an entity is not necessarily
convex, for instance the surface of an octahedron is a polyhedron formed
by the geometric simplicial complex consisting of its faces, edges, and
vertices.

\topic{Contractability}
Intuitively, a set $T\subseteq\R^N$ is \define{contractable} if it can
be continuously shrunk to a single point, while never breaking through
its original boundary.  Technically, $T$ is contractable if there
exists a continuous mapping $\Phi:T\times[0,1]\rightarrow T$ with
$\forall x\in T,\Phi(x,0)=x, \Phi(x,1)=p_0$ for some $p_0\in T$.  One
can show that the choice of $p_0$ is immaterial.

If $T$ consists of 2 distinct points in $\R^1$, no such mapping can
exist because at some time the mapping would have to switch from
mapping a point to itself to mapping the point to the other point.

If the underlying simplicial complex is a graph, if the graph is
disconnected one can similarly show that the set is not contractable.
Similarly, if the graph has a cycle, at any time the cycle will be in
the image of the mapping, so a cyclic graph is not contractable.  

Conversely, if the graph is a tree, then one can contract the graph by
repeatedly contracting the edges leading to leaves.

(Note that we consider contractability of a simplicial complex 
synonymous with the contractability of its geometric realizations.)

Generalizing the contraction of a tree described above, we will obtain
a useful sufficient condition for contractability. (Surprisingly, for
a general simplicial complex $\K$, it is undecidable whether $\K$ is
contractable.)  For $v\in V(\K)$, define\index{$\K\minus v, \K\link v$}
\begin{eqnarray*}
\K\minus v & = & \{X\in\K : v\not\in X\} \\
\K\link v & = & \{X\in\K : v\not\in X, X\cup\{v\}\in\K\}.
\end{eqnarray*}
The first is called \define{$\K$ minus $v$};
the second is called \define{link of $v$ in $\K$}.

\epsFig{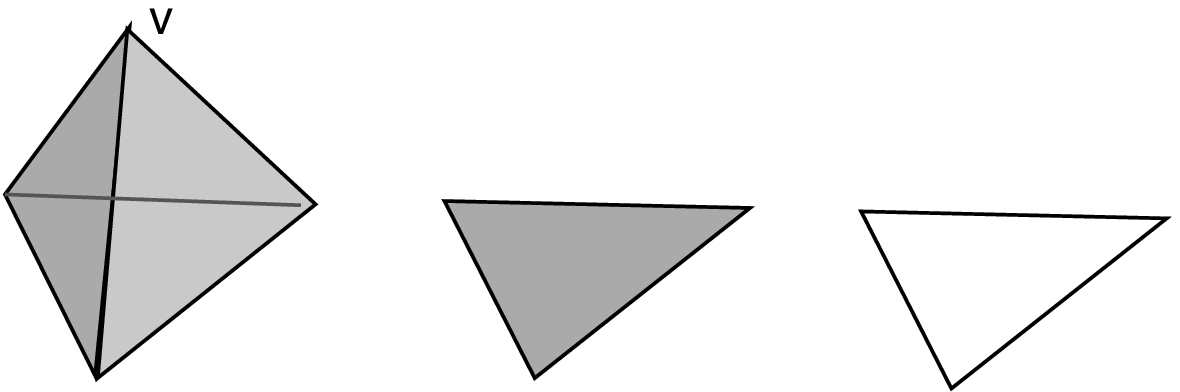}{$\partial \Delta_3$,$\partial \Delta_3 \minus v$,
and $\partial \Delta_3 \link v$.}

Considering the boundary of the 3 dimensional simplex, $\partial \Delta_3$,
which is not contractable,
one sees that $\partial \Delta_3 \minus v \equiv \Delta_2$ is contractable but
$\partial \Delta_3 \link v$, a three node cycle, is not.

\epsFig{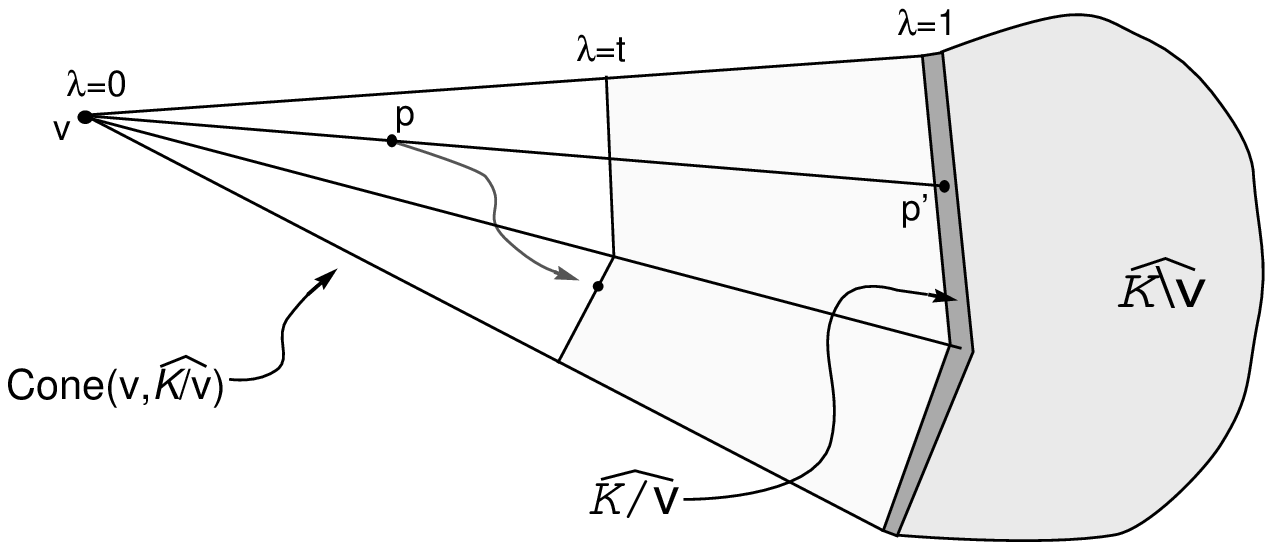}{Contraction of $\widehat{\C}$ onto $\widehat{\K\link v}$.}

\begin{lemma}
If for some $v$, $\K\link v$ and $\K\minus v$ are contractable, then
$\K$ is contractable.
\end{lemma}

\begin{proof}
Let $\C$ denote $\{X\in\K : v\in X\}$, so that
$\widehat{\K} = \widehat{\C}\cup\widehat{\K\minus v}$.

The first step in contracting $\widehat{\K}$ is to use the
contractability of $\K\link v$ to construct a mapping $\Psi$ which
contracts\footnote{
We generalize the notion of contraction to a point in
the natural way to allow contraction to arbitrary contractable
subsets.}
$\widehat{C}$ onto $\widehat{\K\link v}$, leaving
$\widehat{\K\minus v}$ fixed.  Once this is accomplished, all of the
points have been contracted into 
\(\widehat{\K\minus v}\),
so applying the contraction of $\widehat{\K\minus v}$ completes the
contraction of $\widehat{\K}$.

Suppose $\widehat{\K\link v}$ is contracted by $\Phi$ to $p_0$.
Denote a point $p$ in $\widehat{\C}$ by $(p',\lambda)$, where
$p'\in\widehat{\K\link v}$ and $\lambda\in[0,1]$ such that $p=\lambda
p'+(1-\lambda)v$.  (Note that this denotation is continuous and invertible
except at $v$.)

Let $\widehat{\C}_\lambda = \{(p', \lambda):p'\in\K\link v\}$, 
so $\widehat{\C}_1 = \K\link v$ and $\widehat{\C}_0 = \{v\}$.


The contraction can be envisioned as flattening $\widehat{C}$.  At
time $t$, each $\widehat{\C}_\lambda$ for $\lambda<t$ will have been
flattened into $\widehat{\C}_t$, until all of $\widehat{\C}$ is
flattened into $\widehat{\K\link v}$.  Once $\widehat{\C}_\lambda$ is
mapped onto $\widehat{\C}_t$, as $t$ grows, instead of letting the
image of $\widehat{\C}_\lambda$ grow with $\widehat{\C}_t$ (which would
lead to a discontinuity at $v$), we contract it using $\Phi$ to
counteract the growth.
\[\Psi(p,t)=\cases{
\left(\Phi(p',1-\lambda/t),t\right) & if $t \ge \lambda$, and \cr
p & if $t \le \lambda$.
}\]

If $t=0$ or $\lambda=1$ then $\Psi$ is the identity.  If $t=1$ all points
are mapped into $\widehat{\K\link v}$.  We leave it to the reader to verify the
continuity of $\Psi$, remarking only that as $p\rightarrow v$,
$\lambda\rightarrow 0$, so $\Phi(p',\lambda)\rightarrow p_0$,
independent of $p'$.
\end{proof}

\topic{Monotone Functions}
A monotone boolean function $f \not\equiv 1$ gives a simplicial complex
\[\K_f=\left\{S\subseteq\{1,\ldots,n\}:f(x^S)=0\right\}\]
in a natural way, and vice versa.\footnote{
\(\left(x^S\right)_i=\cases{1 & $i\in S$,\cr 0 & $i\not\in S$.}\)}
Also,
\begin{eqnarray*}
\K_\restrict{f}{x_i=0} & = &
\left\{S\subseteq\{1,\ldots,i-1,i+1,\ldots,n\} : S\in\K_f\right\}
=\K_f\minus i, \\
\K_\restrict{f}{x_i=1} & = &
\left\{S\subseteq\{1,\ldots,i-1,i+1,\ldots,n\} : S\cup\{i\}\in\K_f\right\}
=\K_f\link i.
\end{eqnarray*}

By now we may begin to suspect a relation between non-evasiveness and 
contractability.  We prove such a relation in the next lemma.

\begin{lemma}[Kahn-Saks-Sturtevant\index{Theorem, Kahn-Saks-Sturtevant}]
If $f\not\equiv 1$ is non-evasive, then $\K_f$ is contractable.  \end{lemma}
\begin{proof}
Assume $f:\{0,1\}^n\rightarrow\{0,1\}$ is non-evasive, with 
$f\not\equiv 1$.

If $n>1$, then there exists an $i$ such that $\restrict{f}{x_i=0}$ and
$\restrict{f}{x_i=1}$ are non-evasive.  Provided
$\restrict{f}{x_i=1}\not\equiv 1$, we can assume by induction that
$\K_\restrict{f}{x_i=0}$ and $\K_\restrict{f}{x_i=1}$ are
contractable.  By the preceding lemma and remarks, it follows that
$\K_f$ is contractable.

If $n>1$ and $\restrict{f}{x_i=1}\equiv 1$, then
$\K_f=\K_\restrict{f}{x_i=0}$, and $\restrict{f}{x_i=0}$ is
non-evasive, so again by induction $K_f$ is contractible.

Otherwise $n=1$, so $f\equiv 0$ and $\K_f=\{\emptyset,\{1\}\}$, 
which is contractible.

\end{proof}

We have now established a link between evasiveness of monotone
functions and the contractability of the associated simplicial
complex.  In the next lecture, we will use the symmetry properties of
monotone graph properties and some more topology to show in some
cases that the associated complexes are not contractable, and thus
that the original functions are evasive.
\lecture{Fixed Points of Simplicial Maps Show Evasiveness}

In the previous lecture we showed that the simplicial complex
associated with a monotone non-evasive function is contractable.  In
this lecture we present the following argument.

Standard fixed point theorems in topology tell us that a continuous
function mapping a contractable polyhedron into itself has a fixed
point.  On the other hand, the invariance of a monotone function $f$
under a permutation $\pi$ of the inputs implies that the geometric
realization of the permutation maps $\widehat{\K_f}$ into itself, and
thus has a fixed point if $f$ is non-evasive.  For $f$ a monotone
bipartite graph property or a monotone graph property on graphs with a
prime power number of nodes, we characterize the possible fixed point
sets of such mappings to show that if $f$ is non-trivial, no fixed
point can exist, so that $f$ is evasive.

\topic{Fixed Points of Simplicial Mappings}

Suppose $\K$ and $\K'$ are (abstract) simplicial complexes.
Then $\varphi:V(\K)\rightarrow V(\K')$ is a \define{simplicial map}
provided $\forall X\in\K \Rightarrow \varphi(X)\in\K'$, that is,
provided $\varphi$ preserves the property of being in the complex
when applied to sets.  Such a map yields a continuous linear map
$\widehat{\varphi}:\widehat{\K}\rightarrow\widehat{\K'}$ by 
mapping the vertices of $\widehat{\K}$ in correspondence with
$\varphi$, and mapping convex combinations of the vertices
to the corresponding convex combinations of their images:
\[\widehat{\varphi}\left(\sum_{v\in\K}\alpha_v \widehat{v}\right)
= \sum_{v\in\K}\alpha_v \widehat{\varphi(v)}.\]

Note that for $x\in\widehat{\K}$, the representation
\(x=\sum_{v\in\K}\alpha_v \widehat{v} : \sum_v \alpha_v = 1\) is unique.  The set
$\Delta_x = \{v : \alpha_v \neq 0\}$, a simplex of $\K$, is called the
\define{support simplex} of $x$, and is, of course, also unique.

\epsFig{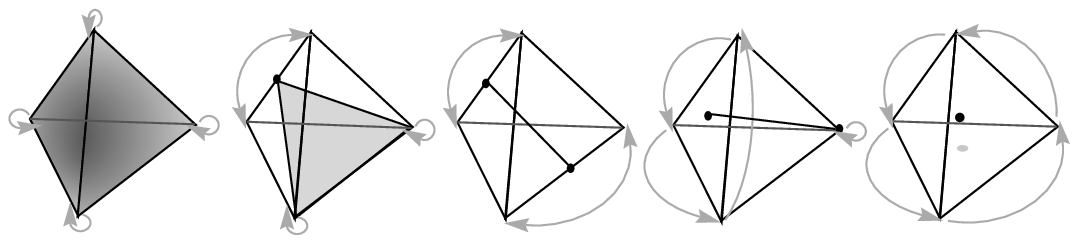}{Fixed Points of Simplicial Maps of $\Delta_3$}

So given a simplicial map $\varphi:\K\rightarrow\K$, which for our
purposes we will assume is one-to-one, what are the fixed points?
Suppose $x$ is a fixed point with support simplex $\Delta$.  Then
$\widehat{\varphi(\Delta)}$ contains $x$, and hence contains
$\widehat{\Delta}$.  Since $\varphi(\Delta)$ and $\Delta$ are the same
size, it follows that $\varphi(\Delta)=\Delta$, that is $\varphi$
permutes the vertices in $\Delta$.  This in turn implies that the
center of gravity\defnote{center of gravity} {The center of gravity of
a face $\widehat{H}$ is $\sum_{v\in H}v/|H|$.} of $\widehat{\Delta}$
is also fixed by $\varphi$.

What are the other fixed points in $\widehat{\Delta}$?  If the
orbits\defnote{orbit} {The orbits, or cycles, of a permutation are the
minimal sets of elements such that the permutation takes no element
out of its set.} of the permutation induced by $\varphi$ on $\Delta$
are $H_1,\ldots,H_k$, then $\widehat{H_i}$ is a face of
$\widehat{\Delta}$, with the center of gravity a fixed point.  Also,
any convex combinations of these centers of gravity is a fixed point.

Conversely, if $x=\sum_{v\in\Delta}\alpha_v \widehat{v}$ is a fixed
point, then 
\(x=\widehat{\varphi}(x)=\sum_{v\in\Delta}\alpha_v \widehat{\varphi(v)}\)
is also a representation of $x$, and because the representation of $x$
in this way is unique, there exists a permutation $\pi$ of the
vertices in $\Delta$ such that 
\(\alpha_{\pi(v)} \widehat{\pi(v)} = \alpha_v \widehat{\varphi(v)}\),
that is, $\varphi(v) = \pi(v)$ and $\alpha_v = \alpha_{\pi(v)}$.  It
follows that $\alpha_v = \alpha_{\varphi(v)}$, so that for each
orbit $H_i$ of $\varphi$ on $\Delta$, we can choose $\beta_i$
so that 
\(u\in H_i\Rightarrow\alpha_u=\beta_i\).
Thus we have 
\[x=\sum_{v\in\Delta} \alpha_v \widehat{v}
=\sum_i \sum_{v\in H_i} \beta_i \widehat{v}
=\sum_i \beta_i |H_i| w_i.\]
In other words, $x$ is a convex combination of the centers of mass of
the faces corresponding to the orbits.

To view this from a more combinatorial perspective, suppose the orbits
of $\varphi$ on the vertices of $\K$ are $H_1,\ldots,H_N$, and assume
that the first $t$ of these are those which are also simplices of
$\K$.  Let $w_i:1\le i\le t$ denote the center of gravity of
$\widehat{H_i}$.  Then each $w_i$ is a fixed point, and any proper
convex combination $x$ of a subset
$\{w_{i_1},\ldots,w_{i_r}\}\subseteq\{w_1,\ldots,w_t\}$ is also a
fixed point, provided only that the $x$ is in fact in $\widehat{\K}$,
that is, provided
\(H_{i_1}\cup\cdots\cup H_{i_r}\in\K\).

In sum, if $\fix(\varphi)$ denotes the fixed points of
$\widehat{\varphi}$, then $\fix(\varphi)=\widehat{\h}$, where
\[\h = \left\{\{i_1,\ldots,i_r\} : H_{i_1}\cup\cdots\cup H_{i_r}\in\K\right\},\]
and the vertices $V(\h)$ of the simplicial complex $\h$ are
$\{1,\ldots,t\}$, with $\widehat{i}$ taken to be the center of
gravity of the face $\widehat{H_i}$.

\topic{Fixed Point Theorems}

Next we give some theorems which give sufficient conditions for the existence
of fixed points, and which characterize some useful properties of the fixed
point sets.  The first is Brouwer's fixed point theorem.

\begin{theorem}[Brouwer\index{Theorem,Brouwer}]
  Any continuous map of a simplex to itself has a fixed point.  
\end{theorem}

An alternate formulation of this theorem follows.
\begin{theorem}
There does not exist a continuous map from a simplex to its boundary
leaving the boundary fixed.
\end{theorem}

If there were a continuous function $f:S\rightarrow S$ with no fixed
point, we could construct a function $g:S\rightarrow\partial S$ leaving
$\partial S$ fixed as follows.  Given $x\in S$, obtain $g(x)$ by
projecting a ray from the point $f(x)$ through the point $x$ to the
boundary $\partial S$.

Note that for any polyhedron $S$, if $\partial S$ is not contractable,
then there can not exist a continuous map from $S$ to its boundary
leaving the boundary fixed.

[Also recall our earlier claim that a set is contractable iff the
cone\defnote{cone}{The cone of a set with a vertex consists of the
convex combinations of $v$ with points in the set.} formed by the set
with an affinely independent point $v$ can be mapped continuously to
the set.]

\begin{theorem}[Lefshetz\index{Theorem, Lefshetz}]
If $\widehat{\K}$ is contractable, then any continuous map from 
$\widehat{\K}$ to itself has a fixed point.
\end{theorem}

\topic{Application to Graph Properties}
We are finally in a position to apply these techniques to show
evasiveness.  Recall the previous result that a function
$f:\{0,1\}^n\rightarrow\{0,1\}$ invariant under some cyclic
permutation of its inputs and with $f(0) \neq f(1)$, is evasive,
provided the number of inputs is prime.  To start, we show that
any non-trivial monotone function invariant under a cyclic
permutation of the inputs is evasive.

\begin{lemma}
Suppose $f:\{0,1\}^n\rightarrow\{0,1\}$ is a monotone, non-trivial
function invariant under a cyclic permutation of its inputs.
Then $f$ is evasive.
\end{lemma}
\begin{proof}
Assume without loss of generality that $f$ is invariant under the
permutation $\varphi(i)=i+1 \mod n$, and assume $f$ is non-evasive.
Consider $\K_f$.  As shown in the previous lecture, $\widehat{\K_f}$
is contractable.  Since $f$ is invariant under $\varphi$, $\varphi$ is
a simplicial map of $\K_f$, so that $\widehat{\varphi}$ has a fixed
point (by Lefshetz' theorem).

As discussed in the beginning of this lecture, the fixed points
correspond to orbits of $\varphi$ contained in $\K_f$.  Since the
only orbit of $\varphi$ is $\{1,2,\ldots,n\}$, this set must be
in $K_f$.  Thus $f(1,1,\ldots,1)=0$, a contradiction.
\end{proof}

The success of this technique hinges on our being able to characterize the
orbits, and hence the fixed point set, of a permutation under which the
function is invariant.  The proof of the next theorem is essentially the same
as the previous, except that the characterization of the orbits is trickier.
Before we give the theorem, we give the Hopf index formula, an extension of
Lefshetz' fixed point theorem which we need for the proof.

\begin{theorem}[Hopf Index Formula\index{Theorem,Hopf}]
For $\varphi$ a simplicial one-to-one mapping of $\K$ a contractable
simplicial complex, the Euler 
characteristic\defnote{Euler characteristic}
{The Euler characteristic $\chi(\widehat{\K})$ of a
polyhedron $\widehat{\K}$ is defined by
\(\chi(\widehat{\K}) = \sum_{x\in\K, x\neq\emptyset}(-1)^{|x|}.\)
(Recall $\mu(f)$.)
The Euler characteristic is invariant under topological deformation,
and is thus useful for classifying topological types.  For instance, a
contractable set has Euler characteristic -1.} 
of the fixed point set $\widehat{\h}$ of $\widehat{\varphi}$ is -1.
\end{theorem}

\begin{theorem}[Yao\index{Theorem, Yao}]\index{AKR conjecture, proof for
bipartite graphs}
\label{bipThm} Non-trivial monotone bipartite graph
properties are evasive.  \end{theorem}
\begin{proof}
Let $f(x_{ij} : i\in U, j\in W)$ be a non-evasive monotone bipartite
graph property.  Let $\varphi$ be a permutation of the edges
corresponding to a cyclic permutation of the vertices of $W$ while
leaving $U$ fixed.  

The fixed point set of $\widehat{\varphi}$ on
$\widehat{\K_f}$ is characterized by:
\[\fix(\varphi)=\widehat{\h},
\{u_{i_1},\ldots,u_{i_r}\}\in\h \Leftrightarrow 
f(x^{\{u_{i_1},\ldots,u_{i_r}\}\times W}) = 0\]
with the vertices of $\widehat{\h}$ being the centers of gravity
of the faces corresponding to the orbits of $\varphi$.

The orbits of $\varphi$ correspond to the nodes of $U$: each orbit
contains all of the edges touching a single node of $U$, so
we identify each vertex of $\h$ with a node of $U$.  Since
$\K_f$ is contractable, there are fixed points, so some edge set
$\{u\}\times W$ has $f(x^{\{u\}\times W}) = 0$.\footnote{Considered as acting
on the complete simplex $\Delta_{|U|\times|W|}$, $\widehat{\varphi}$
necessarily has fixed points.  The question is whether any of these
fixed points are in $\widehat{\K_f}$.}
By the symmetry of $f$, therefore, any choice of $u$ yields
$f(x^{\{u\}\times W}) = 0$.  The sets in $\h$ correspond to 
edge sets $\{u_{i_1},\ldots,u_{i_r}\}\times W$, and again by
symmetry either all or none of these sets for any given $r$
are in $\h$.  By monotonicity, then, $\h$ is characterized by
\[\{u_{i_1},\ldots,u_{i_r}\}\in\h \Leftrightarrow r \le r_0\]
for some $r_0$.  The Euler characteristic of $\h$ is thus
\begin{eqnarray*}
& & -{|U| \choose 1}+{|U| \choose 2}-\cdots+(-1)^{r_0}{|U| \choose r_0} \\
&=& -{|U|-1 \choose 0}-{|U|-1 \choose 1}
+{|U|-1 \choose 1}+{|U|-1 \choose 2}-\cdots \\
&=& -{|U|-1 \choose 0}+(-1)^{r_0}{|U|-1 \choose r_0}.
\end{eqnarray*}
By the Hopf formula, this equals -1, which implies that $r_0=|U|$,
i.e.\ $U\in\h$, so $U\times W\in \K_f$, and $f(x^{U\times W}) = 0$.
Thus $f\equiv 0$.
\end{proof}

One might expect the proof to be simpler for a general graph
property, which is invariant under a larger class of permutations.
Unfortunately, since a general graph has more edges, the orbits
of any given permutation are generally more complicated.  However,
when the number of nodes is a prime power, we can still characterize
the orbits, and thus show evasiveness.  Before we show this, we
give a more general fixed point theorem.
\begin{theorem}
Let $\Gamma$ be a group of mappings of a contractable geometric complex
$\widehat{\K}$ onto itself.  Let $\Gamma_1$ be a normal
subgroup\defnote{normal subgroup}{A subgroup $\Gamma_1$ of $\Gamma$ is
normal if $\forall x\in\Gamma, x\Gamma_1x^{-1} = \Gamma_1$.} of
$\Gamma$ with $|\Gamma_1|=p^k$, for a prime $p$, and with
$\Gamma/\Gamma_1$ cyclic.  Then there exists an $x\in\widehat{\K}$
such that
\(\forall\varphi\in\Gamma,x=\widehat{\varphi}(x)\).
\end{theorem}

Note that we are no longer talking about the fixed points of a single
simplicial mapping, but rather the fixed points of a group of
simplicial mappings.  If the orbits\footnote{The orbits of a
collection of permutations are the minimal sets invariant under every
permutation.} of $\Gamma$ are $H_1,\ldots,H_N$, with the first $t$ of
these in $\K$, and $w_i : 1\le i \le t$ is the center of gravity of
$\widehat{\h_i}$, then $\fix(\Gamma) = \widehat{\h}$, where:
\[\{i_1,\ldots,i_r\}\in\h \Leftrightarrow H_{i_1}\cup\cdots\cup H_{i_r}\in\K,\]
and $V(\h)=\{1,\ldots,t\}$, with $\widehat{i}=w_i$.  

To see this, note first that for any $\varphi\in\Gamma$, an orbit of
$\Gamma$ is expressible as a disjoint union of orbits of $\varphi$, so
any $w_i$ is expressible as a convex combination of centers of mass of
the faces corresponding to orbits of $\varphi$, so $w_i$ is fixed by
$\varphi$.  Thus any convex combination of the $w_i$ is fixed by each $\varphi$.

Conversely, if $x=\sum_{v\in\Delta}\alpha_v \widehat{v}$ is a fixed
point of $\Gamma$, for any $u,w$ in an orbit $H_i$ of $\Gamma$ there
is a $\varphi$ such that $\varphi(u)=w$ and 
\(x=\widehat{\varphi}(x) = \sum_{v\in\Delta}\alpha_v \widehat{\varphi(v)}\),
so the uniqueness of the representation of $x$ implies
\(\alpha_u=\alpha_w\).  Consequently we can choose $\beta_1,\ldots,\beta_t$
such that $\forall i, u\in H_i \Rightarrow \alpha_u = \beta_i$, and
\[x=\sum_{v\in\Delta}\alpha_v \widehat{v}
=\sum_{H_i\subseteq\Delta}\sum_{v\in H_i}\beta_i \widehat{v}
=\sum_{H_i\subseteq\Delta}\beta_i|H_i|w_i.\]

Thus the previous techniques continue to apply when we have a group of
simplicial mappings.  The previous theorem is exactly what we need for
the proof of the next theorem:

\begin{theorem}
\label{primePowThm}
Suppose $f$ is a non-trivial monotone graph property on graphs with a
prime power $p^k$ number of nodes.  Then $f$ is evasive.
\end{theorem}
\begin{proof}
Think of the nodes of our graph $G$ as identified with 
$\GF(p^k)$.\defnote{$\GF(p^k)$}{$\GF(p^k)$ is the Galois field
of order $p^k$.  For $k=1$ this is the field of integers
$0,1,\ldots,p-1$ under arithmetic modulo $p$.  For larger $k$,
it is the field of polynomials over $\GF(p)$ modulo an irreducible
polynomial of order $k$.}
Consider the linear mappings 
\(x\mapsto ax+b:\GF(p^k)\rightarrow\GF(p^k)\)
as a group $\Gamma$.  Let $\Gamma_1$ be the mappings 
\(x\mapsto x+b\), so $|\Gamma_1|=p^k$.  The normality
of $\Gamma_1$ follows from $(((ax+b)+b')-b)/a = x + b/a$.  The factor
group $\Gamma/\Gamma_1$ is isomorphic to the group of mappings
$x\mapsto ax : a\neq 0$, i.e.\ the multiplicative
group\defnote{multiplicative group}{The multiplicative group of a
field is the group formed on the elements other than 0 under
multiplication.} of $\GF(p^k)-\{0\}$, which is known to be cyclic.
Thus the preceding theorem applies, and every action of $\Gamma$ has a
fixed point on $\widehat{\K_f}$.  Since $\Gamma$ is transitive on the
edges, the only orbit of $\Gamma$ consists of all of the edges.  Thus
if $f$ is non-evasive, so that $\K_f$ is contractable and $\Gamma$
has a fixed point, then $\K_f$ must have as an element the set of all
edges, and $f \equiv 0$.
\end{proof}
\lecture{Non-Deterministic and Randomized Decision Trees}

In this lecture we use previous results to show a $\Theta(n^2)$ lower
bound on the decision tree complexity of a general non-trivial
monotone graph property, and we begin discussion of decision 
trees for probabilistic algorithms.

\topic{Near Evasiveness of Monotone Graph Properties}

The key lemma in our study so far of the decision tree complexity of monotone
Boolean functions has been :

{\em If $f:\{0,1\}^n\rightarrow\{0,1\}$ is a non-evasive, monotone function,
with $f(\underline{0})=0$, then $\K_f$ is contractible.}

We also noted that :

{\em If $\K_f$ is contractible then $\chi(\K_f) = -1$
(i.e.\ $\sum_{S:f(S)=0}(-1)^{|S|} = 0$).}

We would like to extend this theory to general (non-monotone)
functions as well.  However, $\K_f$, as defined, is not a simplicial
complex in the case of a non-monotone function $f$.  Although the AKR
conjecture is known to be false for non-monotone functions, if $f$ is
weakly symmetric, $f(\underline{0}) \neq f(\underline{1}) = 1$, and
the number of variables of $f$ is prime, we have shown that $f$ is
evasive.  (Theorem \ref{weakSymThm}).

We have seen that monotone graph properties on graphs with a prime
power number of nodes (theorem \ref{primePowThm}), or on bipartite
graphs (theorem \ref{bipThm}), are evasive.  Next we show that for any
non-trivial monotone graph property, by restricting the property to
some $\Omega(n)$ size subgraph, we can obtain one of these two kinds
of properties.  Since $D(f)$ is at least $D(\restrict{f}{R})$ for any
restriction $R$, this will imply that any non-trivial monotone
property $f$ $D(f)=\Omega(n)$.

\begin{theorem}
Let $f$ be a monotone, non-trivial graph property,
then $D(f) \ge cn^2$ for some positive constant $c$.
\end{theorem}
\begin{proof}
Let $G$ be a graph on $n$ nodes.  Choose a prime $p$ such that $ n/2 <
p < {2n}/3$ (it follows from number theoretic arguments that such a
prime exists).  Let $S$ be a subset of the nodes of $G$ such that
$|S|=p$.  Let $K_S$ denote the complete graph on $S$, with all other
($n-p$) nodes being isolated.  Since $f$ is monotone, and non-trivial,
$f(0) = 0$. Now, there are several cases:
\begin{description}
\item[{\bf Case 1}, $f(K_S) = 1$.]\defnote{$f(G)$}
{$f(G)$ for a graph $G=(E,V)$ is shorthand for $f(X^E)$.}
Let $R$ be the restriction ${x_{i,j} = 0: (i,j) \not\in S}$.
Then $\restrict{f}{R}$ is a monotone, non-trivial graph property
on a graph with a prime number of nodes, and 
\[D(f) \ge D(\restrict{f}{R})={p \choose 2} \ge (n^2-n)/8.\]

\item[{\bf Case 2}, $f(K_S) = 0$.]
In this case $f(K_{V\minus S}) = 0$, since $K_{V\minus S}$ is a complete
graph that is smaller than $K_S$ ($n/2 \le p$), and $f$ is monotone.

Let $H = K_{V\minus S}\cup \{S\times(V\minus S)\}$  (Note the abuse of
notation here, as we are really interested in unordered pairs).

\begin{description}
\item[{\bf Case 2.1}, $f(H) = 1$.]
Let 
\(R=\{x_{i,j} = 0 : i,j \in S\}\cup\{x_{i,j} = 1 : (i,j) \in S\minus V\}\).
Then 
\(\restrict{f}{R}(0) = f(V\minus S) = 0\),
and \(\restrict{f}{R}(1) = f(H) = 1\), and $\restrict{f}{R}$ is a
monotone bipartite graph property on a graph with $p(n-p)$ edges. Thus
$D(f)\ge D(\restrict{f}{R}) = p(n-p) \ge 2 n^2/9$.

\item[{\bf Case 2.2}, $f(H) = 0$.]
Let $R=\{x_{i,j} = 1 : (i,j) \in H\}$.  Then 
$\restrict{f}{R}(0) = f(H) = 0$, $\restrict{f}{R}{1} = f(1)$, and
$\restrict{f}{R}$ is a monotone graph property on the subgraph 
induced by the vertices in $S$, so as in case 1 $D(f)\ge (n^2-n)/8$.
\end{description}
\end{description}
\end{proof}

\topic{Non-Deterministic Decision Trees}

Recall that we defined $D_0(f)$, and $D_1(f)$, and showed that
$D(f) \le D_0(f) D_1(f)$.

\begin{theorem}[Babai or Nisan?\index{Theorem, Babai or Nisan?}]
Suppose $f$ is weakly symmetric (invariant under a transitive group
$\Gamma$), then $D_0(f) D_1(f) \ge n$.
\end{theorem}

\begin{proof}
Recall 
\(D_0(f) = \min \{ k : f=E_1 \wedge E_2\ldots \wedge E_N , E_i =
 x_{i_1}^{\epsilon_{i_1}} \vee \ldots \vee x_{i_k}^{\epsilon_{i_k}}\}\).
Similarly,
\(D_1(f) = \min \{ k : f=F_1 \vee F_2\ldots \vee F_M , F_i =
x_{i_1}^{\epsilon_{i_1}} \wedge \ldots \wedge x_{i_k}^{\epsilon_{i_k}}\}\).
Recall; there must be a variable, $x_i$, that occurs in both $E_1$
and $F_1$ (otherwise, we can force the function value to be $0$, and
$1$, at the same time).  Let $\gamma \in \Gamma$.  Let $E_i^{\gamma}$
be $E_i$ after the action of $\gamma$.  Since $f$ is invariant under
$\Gamma$, we can rewrite$f = E_1^\gamma \ldots E_N^\gamma$.  Therefore
$E_1^\gamma$ must have a variable in common with $F_1$.  

The crucial observation at this point is that for a transitive group
$\Gamma$ of mappings on a set $S$, the quantity
\(q=\#\{\gamma\in\Gamma: \gamma(x) = y\}\)
is independent of $x$ and $y$.  This is because for any $x$, $y$, and
$y'$ we can map 
\(\{\gamma\in\Gamma : \gamma(x) = y\}\) 1-1 into 
\(\{\gamma\in\Gamma : \gamma(x) = y'\}\) by composing any fixed map
$\gamma' : \gamma(y) = y'$ with the maps in the first set.  This shows
independence of $y$ and a similar argument shows independence of $x$.

In fact, we can determine $q$ by the equation (for fixed $x_0$)
\[ q n = \sum_y \#\{\gamma \in \Gamma : \gamma(x_0) = y\} = |\Gamma|,\]
so $q = |\Gamma|/n$.

Returning to the original argument, that $E_1^\gamma$ has a variable
in common with $F_1$ for every $\gamma$ means that every $\gamma$ maps
something from $E_1$ to something in $F_1$, i.e.\ there are $|\Gamma|$
$\gamma$ mapping some $x$ from $E_1$ to some $y$ from $F_1$.  Since
any given pair $x\in E_1$ and $y\in F_1$ (again abusing notation) has
at most $q=|\Gamma|/n$ $\gamma$'s mapping $x$ to $y$, it follows that
there are at least $|\Gamma|/q=n$ pairs $(x,y)$ with $x\in E_1$ and 
$y\in F_1$, i.e.\ $|E_1||F_1| \ge n$.

Recalling that $|E_1| \le D_0(f)$ and $|F_1| \le D_1(f)$ finishes the
argument.
\end{proof}

\topic{Randomized Decision Trees}

The general question in complexity of randomized algorithms is ``Does
the ability to flip a coin add computational power?''  Over the past
twenty years we have learned that the answer is a definitive yes.  
Generally, randomization may give an algorithm the ability to
avoid a few bad computation paths, and thus better its worst case
behavior.

From the decision tree model, there are a number of ways to model
randomized algorithms.  One is that at each node, rather than
definitely querying some variable, we choose which variable to query
randomly according to a probability distribution dependent on the node
and the previous results of random choices.  For a given input the
number of input bits queried is then a random variable, and the
decision tree complexity is the maximum over all inputs of the
expected value of the number of input bits queried.

An alternate model is that the algorithm makes all random choices in
advance, and from that point on is deterministic. In this model an
algorithm for deciding a property is specified as a probability
distribution over all possible decision trees for the property.  For a
given tree $T$, if $\delta(x,T)$ denotes the number of input bits
queried for a given input $x\in\{0,1\}^n$, and $p_T$ denotes the
probability the algorithm choosing $T$, then the decision tree
complexity of the algorithm is 
\[\max_x \sum_T p_T\delta(x,T).\]

This will be made more concrete by the following example.  The example
is due to Saks, Snir, and Wigderson, and gives a tree formula which
has randomized decision tree complexity $o(n)$.  (We have seen
previously that all tree formulae have (deterministic) decision tree
complexity $n$.)

\epsFig{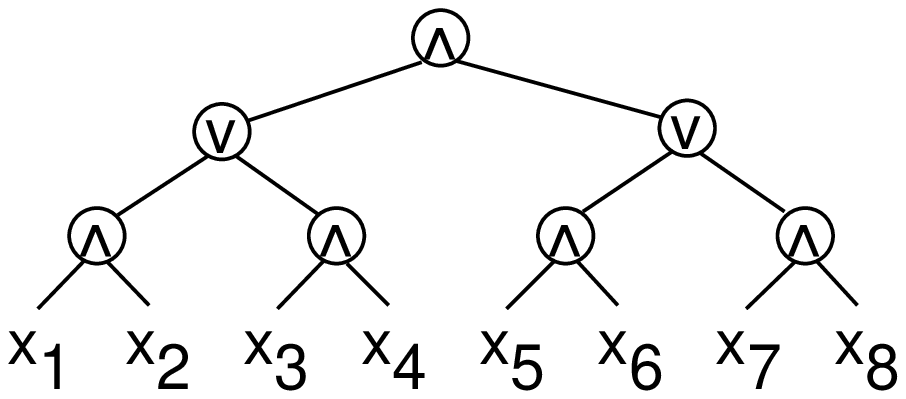}{A Function with Randomized Complexity $o(n)$.}

The function $f_k:\{0,1\}^n\rightarrow\{0,1\}$, $n=2^k$, is a tree
formula. (I.e.\ it is defined by a formula in which each variable
occurs exactly once.)  We may define $f_k$ inductively by
\begin{eqnarray*}
f_0(x_1) & = & x_1 \\
f_{k+1}(x_1,...,x_n) & = &
\cases{ f_k(x_1,...,x_{n/2}) \wedge f_k(x_{n/2+1},...,x_n) & $k$ odd, \cr
	f_k(x_1,...,x_{n/2}) \vee f_k(x_{n/2+1},...,x_n) & $k$ even.}
\end{eqnarray*}
That is, we take a balanced binary tree and construct a formula by labeling
the leaves with the variables and labeling internal nodes of the tree
alternately with ``and'' and ``or'' gates as we go up the tree.

We saw previously that tree formulae are evasive.  The evasiveness of $f$
also follows by the weak symmetry of $f$ and the fact that the number of 
variables is a prime power.  Our task is to construct a randomized algorithm
for $f$ with expected decision tree complexity $o(n)$.

First, for convenience, we get rid of the asymmetry at different levels by 
replacing the and-gates and or-gates by nand-gates (negated and-gates).  
Specifically, we use 
\((f_1 \vee f_2) \wedge (f_3 \vee f_4) = 
(f_1 \overline{\wedge} f_2) \overline{\wedge} (f_3 \overline{\wedge}
f_4)\) to replace all gates except possibly the gate at the root with
nand-gates.  If the gate at the root does not become a nand-gate, it
is an and-gate, and we simply negate it.  This complements $f$, but
doesn't change the complexity of the function.

Now that we have nand-gates at all the nodes, consider the evaluation
of the function.  If for some nand-gate, we know one input is 0, we
know the gate outputs 1, independent of the other input.  Thus our
randomized strategy will be to start at the root, choose one of the
two inputs uniformly at random to evaluate, and recursively evaluate it.
If it returns 0 we return 1, otherwise we evaluate the other input
recursively, returning 1 if the other input returns 0, and 1 otherwise.

Let $a_k$ denote the expected number of variables checked to compute
$f$ if $f(x)=0$, and let $b_k$ denote the expected number if $f(x)=1$.

If $f(x)=0$, then both inputs must be evaluated and are 1.
If $f(x)=1$, then either both inputs are 0, in which case we
definitely only evaluate one input, or one input is 0, in which
case we have at least a one in two chance to evaluate only one
input. This yields
\begin{eqnarray*}
b_k & = & 2 a_{k-1}, \\
a_k & \le & 
\max\{b_{k-1}, \frac{1}{2}b_{k-1} + \frac{1}{2}(a_{k-1}+b_{k-1})\} \\
& = &  \frac{1}{2}b_{k-1} + \frac{1}{2}(a_{k-1}+b_{k-1}).
\end{eqnarray*}
We can write this as
\begin{eqnarray*}
\left(\begin{array}{c} a_0 \\ b_0\end{array}\right) & = &
\left(\begin{array}{c} 1 \\ 1\end{array}\right), \\
\left(\begin{array}{c} a_k \\ b_k\end{array}\right) & = &
\left(\begin{array}{cc} \frac{1}{2} & 1 \\ 2 & 0\end{array}\right)
\left(\begin{array}{c} a_{k-1} \\ b_{k-1}\end{array}\right) \\
& = &
\left(\begin{array}{cc} \frac{1}{2} & 1 \\ 2 & 0\end{array}\right)^k
\left(\begin{array}{c} 1 \\ 1\end{array}\right).
\end{eqnarray*}
To estimate $a_k$ and $b_k$ from such a recurrence relation, we can
examine the eigenvalues of the matrix.  The action of the matrix on an
eigenvector (by definition) is just to stretch the vector by the
corresponding eigenvalue.  Thus on repeated application of the matrix,
the norm of the eigenvector with largest eigenvalue will grow most
rapidly.  For a vector which is not an eigenvector, we can consider it
as a convex combination of eigenvectors and similarly show that with
repeated application of the matrix, its norm grows no faster than that
of the eigenvector with largest eigenvalue.  These considerations show
that if there is a single largest eigenvector $\lambda$, then after
$k$ applications of the matrix the resulting vector has norm
$\lambda^k+o(\lambda^k)$ times the norm of the original vector.  Thus
$a_k$ and $b_k$ are approximately $\lambda^k$.  For the above matrix,
the eigenvalues are $\frac{1}{4}(1\pm \sqrt{33})$, so that 
$a_k, b_k \approx \left(\frac{(1\pm \sqrt{33})}{4}\right)^k$.

For our randomized algorithm, $k=\log_2 n$, so we have
\[a_k, b_k \approx \left(\frac{(1 + \sqrt{33})}{4}\right)^{\log_2 n}
= n^{\log_2 \left(\frac{(1 + \sqrt{33})}{4}\right)}
\approx n^{0.754} = o(n),\] and we are done.  (See the end of this
lecture for details of the calculation.)

One might suspect that one could improve this algorithm by sampling
the inputs randomly and using the result to bias the choice of
which input to evaluate first towards the input with more 1's in
the subtree.  It turns out this doesn't help; in fact the above algorithm
is essentially optimal.  (Although we don't give the proof.)

In the next lecture we will begin to study the probabilistic version of the
AKR-conjecture\index{AKR conjecture, probabilistic}, which is that for a graph
of $n$ nodes the expected decision tree complexity of a randomized algorithm
for a non-trivial graph property is $\Theta(n^2)$.  A lower bound of $n$
follows from $D_R(f) \ge D_0(f) D_1(f) \ge n$.  Yao\index{Yao} improved this
lower bound to $n\log n$ and introduced some techniques which we will study.
Valerie King\index{King} improved the bound to $\Theta(n^{5/4})$ in her
thesis, and subsequently Hajnal\index{Hajnal} improved the bound to
$\Theta(n^{4/3})$.

One question which arises is the complexity of probabilistic
algorithms which are allowed a small probability of error.  (Called a
\define{Las Vegas algorithm}.)  One can gain a little bit, for instance
consider the majority function, which is 1 provided a majority of its
inputs are 1.  By random sampling half the inputs, say, one can
compute the majority function with a small probability of error.
On the other hand, there is a lower bound on the complexity of
$\sqrt{D(f)}$, which leaves a large gap.

\subtopic{Details of Calculating $a_k$ and $b_k$}
Following are the calculations of $a_k$ and $b_k$ in more detail; they
may be skipped by anyone familiar with linear algebra.  Let
\(M=\left(\begin{array}{cc}
\frac{1}{2} & 1 \\ 2 & 0\end{array}\right)\).  The \define{eigenvalues} of $M$
are the values $\lambda\neq 0$ such that there exists a corresponding
\define{eigenvector} $x$ with $Mx = \lambda x$.  Rewriting this as
$(M-\lambda I) x = 0$, we see that the eigenvalues are those values
for which $M-\lambda I$ is singular, i.e.\ has determinant $|M-\lambda
I|=0$.

\begin{eqnarray*}
\left|\begin{array}{cc}
 \frac{1}{2}-\lambda & 1 \\ 2 & -\lambda\end{array}\right|
 & = & 0, \\
\lambda^2 - \frac{1}{2}\lambda - 2 & = & 0, \\
\lambda & = & \frac{1}{4}(1\pm \sqrt{33}).
\end{eqnarray*}

Once we have the eigenvalues $\lambda_1$ and $\lambda_2$, let $x_1$
and $x_2$ be corresponding eigenvectors, and define the matrices
\begin{eqnarray*}
D & = & 
\left(\begin{array}{cc} \lambda_1 & 0 \\ 0 & \lambda_2\end{array}\right), \\
X & = & \left(\begin{array}{cc} x_1 & x_2\end{array}\right). \\
\end{eqnarray*}

It is easy to verify that $M X = X D$, so $M = X D X^{-1}$, 
$M^2=XD^2 X^{-1}$, ..., and $M^k = X D^k X^{-1}$.  Since
\[D^k = 
\left(\begin{array}{cc} \lambda_1^k & 0 \\ 0 & \lambda_2^k\end{array}\right),\]
we have
\[\left(\begin{array}{c} a_k \\ b_k\end{array}\right)
= M^k \left(\begin{array}{c} 1 \\ 1\end{array}\right)
= X D^k X^{-1} \left(\begin{array}{c} 1 \\ 1\end{array}\right).\]
After one determines the eigenvectors $x_1$ and $x_2$, it is 
an easy matter to complete this and give a closed form for $a_k$
and $b_k$.

\vspace{1in}
(Notes by Sigal Ar and Neal Young.)    
\lecture{Lower Bounds on Randomized Decision Trees}

We begin this lecture with a proof of a basic result, Farkas' lemma.
The lemma gives a necessary and sufficient condition for the existence
of a solution to a set of linear inequalities.  We then discuss (and
prove) von Neumann's min-max theorem.  The theorem gives some insight
into the advantages of randomization.  We then present some techniques
developed by Yao applying the min-max theorem to give lower bounds on
randomized decision tree complexity.

\topic{Farkas' Lemma}
For a system of linear equalities, everyone knows necessary and
sufficient conditions for solvability.  Farkas' lemma is the analogue
for systems of linear inequalities.  Consider the problem ``Does a
system of linear inequalities
\begin{eqnarray}
\label{primal}
\sum_{j=1}^m a_{ij}x_j & \le & b_i \ \ (i=1,...,n)
\end{eqnarray}
have a solution?''  Roughly, we expect this problem to be in NP, since
we can exhibit an easy proof (an assignment of $x$) if it is.  Farkas'
lemma implies that it is also in co-NP, that is, if it is not solvable
there is also an easy proof that it isn't.

If the system is not solvable, we will prove it by exhibiting $\lambda_i$
satisfying 
\(\sum_i \lambda_i a_{ij} = 0\ (j=1,...,m)\),
\(\sum_i \lambda_i b_i < 0\), and
\(\lambda_i \ge 0\ (i=1,...,n)\).  This provides a proof that
no $x$ satisfies (\ref{primal}), because if such an $x$ existed we
would have
\[0 = \sum_j x_j \sum_i \lambda_i a_{ij}
= \sum_i \lambda_i \sum_j a_{ij} x_j
\le \sum_i \lambda_i b_i < 0.\]

\begin{lemma}[Farkas]\index{Theorem, Farkas}
For any $a_{ij}$ and $b_i$, ($j=1,...,m$, $i=1,...,n$)
\begin{equation}
\sum_j a_{ij}x_j \le b_i\ \ (i=1,...,n) \label{primalEqn}
\end{equation}
has a solution $x$ if and only if
\begin{eqnarray}
\sum_i \lambda_i a_{ij} & = & 0 \ \ (j=1,...,m) \nonumber\\
\sum_i \lambda_i b_i & < & 0 \label{dualEqn} \\
\lambda_i & \ge & 0  \ \ (i=1,...,n) \nonumber
\end{eqnarray}
has no solution $\lambda$.
\end{lemma}
\begin{proof}
Consider the vectors $(a_{i,1}, ..., a_{i,m}, b_i), (i=1,...,n)$.
Consider the cone of these vectors.  (\ref{dualEqn}) says exactly
that $y^*=(0,...,0,-1)$ is in this cone.  Suppose that (\ref{dualEqn})
has no solution, so $y^*$ is not in the cone.  Then there exists a
separating hyperplane $H=\{y : \sum_j y_j h_j = 0\}$ such that $y^*$
is on one side of $H$ and the cone is on the other, i.e.\
\(\sum_j y^*_j h_j < 0\) and
\(a_{i1}h_1 + \cdots + a_{in}h_n + b_i h_{n+1} \ge 0\ \ (i=1,...,m)\).
The first condition says that $h_{n+1} \ge 0$, and the second
says that \(\sum_j a_{ij} h_j/h_{n+1} \ge -b_i\ \ (i=1,...,n)\).
Thus letting $x_j = -\frac{h_j}{h_{n+1}}$, we have a solution
to $\ref{primalEqn}$.
\end{proof}

\topic{Von Neumann's Min-Max Theorem}
Recall our second characterization of a randomized algorithm via
decision trees.  For a function $f:\{0,1\}^n\rightarrow\{0,1\}$, on an
input $x$, an algorithm $\A$ chooses a deterministic algorithm for
$f$ with decision tree $T$ according to some probability distribution
$p$ (independent of $x$!), and then runs the deterministic algorithm
on $x$.

On a given input $x$, having chosen a particular tree $T$, $\A$
(deterministically) takes some complexity $\delta(T,x)$ to compute $f(x)$.
The expected complexity for $\A$ on $x$ is given by
\(\sum_T p_T\delta(T,x)\).  We are interested in the worst case
expected complexity for $\A$ (j.e.\ the adversary chooses $x$ to maximize
the complexity): \(\max_x \sum_T p_T\delta(T,x)\).  Finally, if $\A$ is
optimal, it minimizes this worst-case complexity, and thus takes complexity
\[D_R(f) = \min_p \max_x \sum_T p_T\delta(T,x).\]

We can view this process as a game, in which we choose $p$
(determining $\A$), and then the adversary, knowing our choice,
chooses $x$.  To play the game, we run $\A$ on $x$.  Our goal is to
minimize the expected complexity; the adversary's goal is to maximize it.

For this situation (called a \define{zero-sum game}, since we lose
exactly what our opponent gains), there is a general theorem, von
Neumann's min-max theorem.  We have a game, defined by a matrix $M$,
for two players --- the row player and the column player.  The game is
played as follows.  The column player chooses a column $c$ and the row
player chooses a row $r$.  The column player then pays $M_{rc}$ to
the row player.

The column player wants to minimize $M_{rc}$, and the row player
wishes to maximize it.

\epsFig{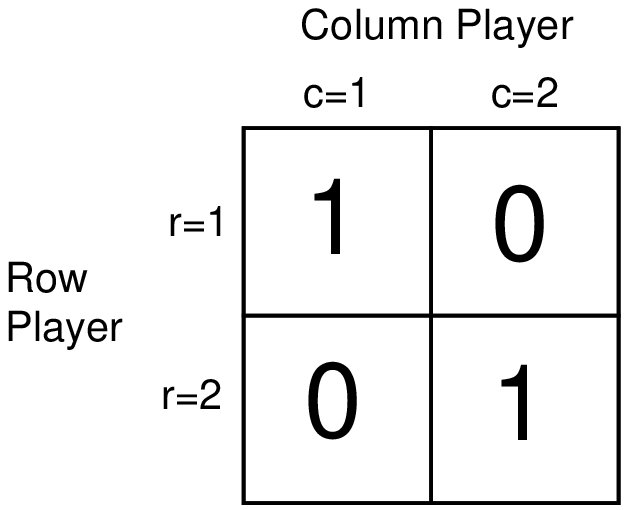}{A Simple Zero-Sum Game}

To make this concrete, assume we are playing a game (see figure
\ref{identityGame}) where each player chooses either 1 or 2.  If we
choose the same as the adversary, we pay her 1.  Otherwise we pay
nothing.  What should our strategy be?  Suppose our strategy is to
choose 1.  After playing the game several times, with the adversary
beating us every time, we begin to suspect that the adversary knows
our strategy and has used that knowledge to beat us.  What can we do,
given that the adversary may know our strategy and use that
information to try to beat us?

Von Neumman's key observation is that we can use randomization to
negate the adversary's advantage in knowing our strategy.  For our
simple game, if our strategy is to choose 1 or 2 randomly, each with
probability 1/2, then no matter what strategy the adversary picks, we
have an expected loss of at most 1/2.

More generally, in any zero-sum game we have a randomized (also called
a {\em mixed}) strategy which negates the adversary's advantage in
knowing our strategy.  To make precise the notion of ``negating the
advantage,'' we consider turning the tables, so that she chooses her
strategy first, and then we choose ours.  Then provided we choose the
randomized strategy that negates her advantage in the first situation,
and she chooses the randomized strategy which negates our advantage in
the second situation, we will expect to do just as well in the first
situation as the second.  Formally,

\begin{theorem}[Von Neumann's min-max theorem\index{Theorem, Von Neumann}]

For any zero-sum game $M$,
\[ \min_p \max_q q\transpose M p = \max_q \min_p q\transpose M p .\]
\end{theorem}
(Note that $p$ and $q$ range over all probability distributions of
columns and rows, respectively.)

We start with some observations.
In the min-max theorem, in the inner $\max$ and $\min$, randomization
is not important.  That is, (if $e_j$ denotes the $j$th unit vector in
a vector space implicit in the context)
\[\forall p, \max_q p\transpose M p = \max_j e_j\transpose M p,\]
and similarly for the inner term on the right hand side.  This is
because, for a fixed $p$, $q\transpose M p$ is a linear function of
$q$, and thus is maximized at one of the vertices $e_j$ of the
probability space.  

\begin{proof}
Obviously,
\[\forall q_0,p_0,
\max_q q \transpose M p_0
\ge q_0\transpose M p_0
\ge \min_p q_0\transpose M p.\]
Thus
 \[\min_{p_0} \max_q q \transpose M p_0 
\ge \max_{q_0} \min_p  q_0\transpose M p.\]

The other direction is not so easy.  We suppose that $\exists t :$
\[\forall p: (p \ge 0, \sum_j p_j = 1)
\ \max_j e_j\transpose M p \ge t\] 
and we want to show that
\[\exists q : q\ge 0, \sum_j q_j = 1,  \forall i, q\transpose M e_i \ge t.\]

This will follow almost immediately from Farkas' lemma.  

Suppose that
\(\not\exists q : 
q\ge 0, \sum_j q_j = 1, \forall i, q\transpose M e_i\ge t\).
Farkas' lemma implies that there exists $\lambda=(\lambda_1,..,\lambda_n)$,
$\mu=(\mu_1,...,\mu_m)$, and $\alpha$ such that:
\begin{eqnarray}
\label{sixB}
\lambda\transpose M + \mu +
\alpha\left(\begin{array}{c} 1 \\ \vdots \\ 1 \end{array} \right)
& = &
\left(\begin{array}{c} 0 \\ \vdots \\ 0 \end{array} \right), \\
\label{sixC}
\sum_i \lambda_i t + \alpha & > & 0, \\
\lambda_i, \mu_j &\ge & 0.  \nonumber
\end{eqnarray}
Note that $\alpha$ is not constrained to be non-negative and that in
(\ref{sixC}) the expression is constrained to be positive, rather than
negative.  These are essentially trivial variations from the standard
form of Farkas' lemma.  The first is because the inequality
corresponding to $\alpha$ is in fact an equality, and the second is
because the unsatisfiable constraints are of the form $\cdots \ge t$,
rather than $\cdots \le t$.

Letting $p =\lambda/\sum_i \lambda_i$, (\ref{sixB}) and (\ref{sixC}) imply
\[ p\transpose M
\le -\frac{\alpha}{\sum_i\lambda_i}
\left(\begin{array}{c} 1 \\ \vdots \\ 1 \end{array} \right)
\le t
\left(\begin{array}{c} 1 \\ \vdots \\ 1 \end{array} \right),\]
which contradicts our assumption.
\end{proof}

\topic{Lower Bounds}
The discussion before the proof, in our original context, means that
once we choose $p$, determining $\A$, the adversary can choose a
specific input $x$, rather than a distribution of inputs, to give the
worst case expected behavior for $\A$.  Alternatively, if the
adversary chooses an input distribution first, then we can do our best
by subsequently choosing the best deterministic algorithm, rather than
a randomized algorithm, for this input distribution.  Thus the min-max
theorem implies that if the adversary can choose a distribution for which
she can force any deterministic algorithm to have an expected complexity of
at least $t$, then for every randomized algorithm there is an input
such that the expected complexity of the algorithm on that input is at least
$t$.

To show a lower bound on the decision tree complexity of any
randomized algorithm, then, it suffices to show the same bound on the
complexity of all deterministic algorithms for some fixed input
distribution.  This is the technique we will use.

As a starting point, recall that $D_R(f)$ denotes the minimum expected
decision tree complexity of a randomized algorithm for $f$.  An almost
trivial observation is
\begin{equation}
\label{sixD}
D_R(f) \ge \max\{D_0(f), D_1(f)\}.
\end{equation}
This follows because when $f(x)=i$, {\em any} algorithm must look at
at least $D_i(f)$ bits of $x$ before it can be sure $f(x)=i$.  Last
lecture we showed that for (non-trivial) weakly symmetric $f$,
$D_0(f)D_1(f)\ge n$.  Thus $D_R(f) \ge \sqrt{n}$.

Next we give a lemma which uses the min-max theorem to give
a better lower bound for certain types of functions.

\begin{lemma}[Yao]
\label{sixYaoLemma}
Let $f:\{0,1\}^n \rightarrow \{0,1\}.$  Suppose we can partition
$\{1,...,n\}$ into $S_1,...,S_{r}$ so that $|S_i| \ge t$ and
\[ f(x) = 
\cases{0 & if $\forall i \left|\{j:x_j = 0\}\cap S_i\right| \ge t$, \cr
1 & if $\exists i \left|\{j:x_j = 0\}\cap S_i\right| = 0$.}\]

Then $D_R(f)\ge \Omega\left(\frac{n}{t}\right)$.
\end{lemma}
(Note that the condition on $f$ leaves some values of $f$ unspecified.
Also note that this bound doesn't follow immediately from
(\ref{sixD}).)
\begin{proof}
We give an input distribution on $x$ and give a lower bound
on the expected complexity of any deterministic algorithm on
this input distribution.

To generate $x$, choose $t$ indices $j$ uniformly at random from each
$S_i$ and set $x_j=0$.  Set the remaining $x_{j'}=1$.  Then $f(x)=0$,
and for an algorithm to verify this it must query an $x_j$ with the
value 0 from each $S_i$.  How many queries must the algorithm expect
to make before finding such an $x_j$ in a given $S_i$?  Since the $t$
$x_j$ with value 0 were chosen randomly, we can view the algorithm as
sampling randomly (without replacement) from a set of size $k$ until
it finds one of the $t$ $x_j$ chosen to be 0.  The expected time for
this is $\Omega(k/t)$.  By linearity of expectation, the expected time
to find a 0 in each $S_i$ is thus at least
$\frac{n}{k}\Omega\left(\frac{k}{t}\right) = \Omega(n/k)$.
\end{proof}

We will apply this lemma to lower bound the randomized complexity of monotone
bipartite graph properties on $2n$ nodes ($n$ in each part).  The complexity
of such properties is not quite solved.  The trivial bound $D_R(f)\ge
\sqrt{{\rm \#inputs}}$ gives a lower bound of $n$.  This was first improved to
$n\log n$ by Yao\index{Yao}, who introduced the general techniques for the
problem.  Valerie King\index{King} improved the bound to $n^{5/4}$, and
subsequently Hajnal\index{Hajnal} improved the bound to $n^{4/3}$.

\subtopic{Graph Packing}
Our problem is closely related to the problem of \define{graph
packing}.  Two graphs $G_1$ and $G_2$ on $n$ nodes can be packed if
(possibly after relabeling the vertices) the edge sets of
the graphs don't overlap.  If $G_1$ and $G_2$ pack, then 
$G_1\subseteq \overline{G_2}$.

Suppose $G_1$ is a minimal $G$ in a graph property
\(\p_f =\{G : f(G) = 1\}\),
and $G_2$ is a minimal $G$ such that $\overline{G}\in\p$.
Observe that $G_1$ and $G_2$ don't pack.  Furthermore, if
any two $G_1$ and $G_2$ don't pack, we have
\(\left|E(G_1)\right| \left|E(G_2)\right| \ge n^2\) by essentially
the same argument that showed $D(f)\ge D_0(f) D_1(f)$.
More interestingly, one can show
\[ d_{\max}(G_1) d_{\max}(G_2) \ge n/2. \]  An instance of this
is that if a graph $G$ has $d_{\min}(G) > n/2$, then there exists a
hamiltonian cycle\index{Hamiltonian cycle} $C$.  That is, if
$d_{\max}(\overline{G}) < n/2$, then $\overline{G}$ and any cycle $C$ pack.

\subtopic{Yao's $d_{\max}/\overline{d}$ Lemma}

Next we will in some sense specialize lemma \ref{sixYaoLemma} to the
case of a monotone bipartite graph property.  We will choose a minimal
graph $G$ for the property, and show how to construct a containing
graph $G'$ such that we can partition a large subset of the edges of
$G'$ into disjoint sets such that if we remove edges from the subsets,
the property holds if we leave any set untouched, but fails if we
remove at least a small number from each set.  Lemma \ref{sixYaoLemma}
will apply to show essentially that any randomized algorithm has to
check many edges in each set.

\begin{lemma}[Yao]\index{Theorem, Yao}
Let $f$ be a (non-trivial) monotone bipartite graph property on
bipartite graphs with $2n$ nodes, the two parts $U$ and $W$ each
having $n$ nodes, and let $\p=\{G : f(G) = 1\}$.

If $G$ is a graph in $\p$ with minimum $d^{U}_{\max}(G)$
(i.e.\ minimum maximum degree over vertices in $U$)
and $\overline{d}^{U}(G)$ is the average degree of 
vertices of $G$ in $U$, then
\[ D_R(f) \ge \Omega(1)\frac{d^{U}_{\max}(G)}{\overline{d}^{U}(G)}. \]
\end{lemma}
\begin{proof}
Of the $G\in\p$ with minimum $d_{\max}$, choose one with the fewest
number of maximum degree vertices, so that if any graph has fewer
vertices of degree $d_{\max}$ (and no higher degree vertices) it is
not in $\p$.


Assume $d_{\max} \ge 4\overline{d}$, otherwise the bound is trivial.
Assume also that the vertices are labeled so that vertex $0$ is a
maximal degree vertex in $U$, and vertices $1,...,n/2$ are in $U$ and
have degree at most $h=2\overline{d}$. 

We form the containing graph $G'$ by adding edges from vertex $i$ to
the neighbors of vertex $0$ and vertex $i+1$.  $G'$ has two essential
properties.  First, if we delete any $2h+1$ edges out of each vertex
$0,1,...,n/2$, we destroy the property $\p$, because we reduce the
number of vertices in $U$ with degree $d_{\max}$ by 1.  

Second, if for some $i$ (possibly 0), for each vertex
$j=0,1,...,i-1,i+1,...,n/2$, we delete the edges from $j$ into
$\Gamma_0 - \Gamma_i - \Gamma_{i+1}$, we preserve $\p$.  This
is because we can permute the vertices of $U$ so that the permuted
graph contains the original graph $G$.  We do this as follows: shift
vertex 0 to vertex 1, vertex 1 to vertex 2, ..., and vertex $i$ to
vertex 0.

To apply lemma \ref{sixYaoLemma}, we define
\(S_i=\{i\}\times\left(\Gamma_0-\Gamma_i-\Gamma_{i+1}\right)\)
(for $i=1,...,n/2$), 
\(S_0=\{0\}\times\left(\Gamma_0-\Gamma_1\right)\), and $S=\cup_i S_i$.

Then if we obtain $f'$ by restricting the domain of $f$ to graphs
which agree with $G'$ on edges not in $S$, $f'$ is a function of
\(|S|\ge n\left(d_{\max}-2h\right)\) variables.  If we start with $G'$ and
within each partition $S_i$ delete $2h+1$ edges, $f'$ becomes 0, but
if we start with $G'$ and delete edges within $S$ leaving at least one
partition complete the function stays 1.  Thus lemma \ref{sixYaoLemma}
implies that
\[D_R(f)\ge D_R(f')
\ge \Omega\left(\frac{n\left(d_{\max}-2h\right)}{2h+1}\right)
\ge \Omega\left(n\frac{d_{\max}}{\overline{d}}\right).\]
\end{proof}

\lecture{Randomized Decision Tree Complexity, continued}

In this lecture we continue giving lower bounds on randomized
decision tree complexity, combining the various bounds we have
developed to show a lower bound of $\Omega(n^{5/4})$ on the
randomized decision tree complexity of any non-trivial,
monotone, bipartite graph property.

The general method is to choose both a minimal graph with the
property and a graph whose complement is minimal in the complementary
property, and then to use the fact that the two graphs don't pack
to get constraints on the maximum and average degrees of the two
graphs, and finally to apply Yao's lemma from last lecture to
bound the randomized complexity via the constraints on the degrees.

\topic{More Graph Packing}
Recall that graphs $G_1$ and $G_2$ pack if after
some relabeling of the nodes of $G_1$ the two graphs have no common
edges.  That is, $\exists G_1', G_1' \cong G_1, G_1' \subseteq
\overline{G_2}$.

Recall that $d^U_{\max}(G)$, for a bipartite graph $G=(U,W,E)$,
denotes the maximum degree of a vertex in $U$ in $G$, and
$\overline{d}(G)$ denotes $|E|/n$, the average degree of a vertex in
$G$.  For this lecture, we will restrict our attention to bipartite
graphs $G_1=(U,W,E_1)$ and $G_2=(U,W,E_2)$ with equal size parts,
i.e.\ $|U|=|W|=n$, so that the average degree in each part is the
same.

We will show the following lemma.
\begin{lemma}[Bollobas-Eldridge]\index{Theorem,Bollobas-Eldridge}
\label{sevenPackLemma} 
If
\[d^U_{\max}(G_1)d^W_{\max}(G_2) + d^U_{\max}(G_2)d^W_{\max}(G_1) \le n\]
then $G_1$ and $G_2$ pack.
\end{lemma}

There is also a non-bipartite version of this lemma:
\begin{lemma}
For two graphs $G_1$ and $G_2$, if
\[d_{\max}(G_1)d_{\max}(G_2) \le n/2\]
then $G_1$ and $G_2$ pack.
\end{lemma}

The proof of the second lemma, which we omit, is similar to that of
the first lemma, which we give.  While there is no gap in the $n/2$
term; the conjecture is that
\((d_{\max}(G_1)+1)(d_{\max}(G_2)+1) < n\) also gaurantees packing.

\begin{proof}
\epsFig{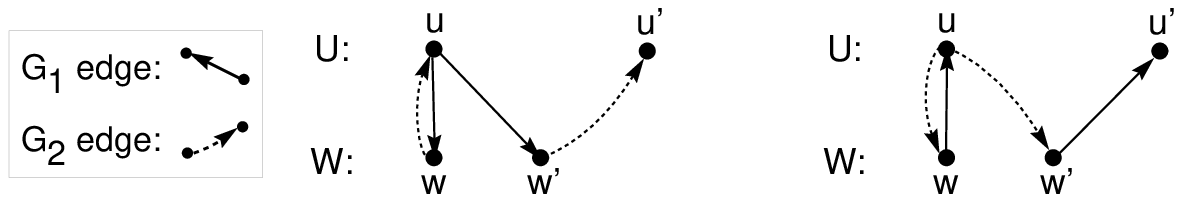}{Swapping the Labels of $u$ and $u'$: Two Cases.}
(Lemma \ref{sevenPackLemma}.)  Suppose $G_1$ and $G_2$ don't
pack, and we have relabeled the vertices of $G_1$ so as to minimize
the number of overlapping edges.  There is some overlapping edge
$(u,w)$.  Consider swapping the labels of $u$ and a vertex $u'\in U$ 
in $G_1$.  With the current labeling, there is at least one overlapping
edge out of $u$ and $u'$, so after the swap this must also be the case.

This entails that either an edge $(u,w') \in E_1$ will be mapped on
to an edge $(u',w') \in E_2$ by the swap, or that an edge $(u',w')
\in E_1$ will be mapped onto an edge $(u, w') \in E_2$ by the
swap.  This means that every $u' \neq u$ is reachable from $u$ either
by following an edge in $G_1$ and then an edge in $G_2$ or by following
an edge in $G_2$ and then an edge in $G_1$.  

There are at most $d^U_{\max}(G_1) d^W_{\max}(G_2) - 1$ paths of the first
kind, (one of the candidates leads back to $u$), and similarly at most 
$d^U_{\max}(G_2) d^W_{\max}(G_1) - 1$ of the second kind.  Thus
\[d^U_{\max}(G_1)d^W_{\max}(G_2) + d^U_{\max}(G_2)d^W_{\max}(G_1) \ge n + 1.\]
\end{proof}

Graph packing captures many graph theoretic notions, for instance if $G_1$ is
a $K_d$ (a graph with a $d$-clique and $n-d$ isolated nodes), and $G_2$
consists of $d$ $n/d$ cliques, then $G_1$ and $G_2$ pack, and this is
equivalent to saying the vertices of $G_1$ are
$d$-colorable\defnote{$d$-colorable}{The vertices of a graph are $d$-colorable
if one can color them with $d$ colors so no edge touches two vertices of the
same color.  The edges of a graph are $d$-colorable if one can color them with
$d$-colors so no vertex touches two edges of the same color.} with each color
coloring $n/d$ nodes.  In fact it can be proved\footnote{Hajnal\index{Hajnal}
and Stever\'edi\index{Stever\'edi}?} that if $G_1$ is of maximal degree $d-1$,
then $G_1$ and $G_2$ pack.

On the other hand, if $G_1$ instead consisted of a $d+1$-clique and
$n-d-1$ isolated nodes, then $G_1$ and $G_2$ would not pack, since
$G_1$ would not be $d$-colorable.  Since the product of the maximum
degrees for this pair of graphs is $d(n/d - 1) = n - d$, which for
$d=n/2$ is $n/2$, this gives a tight lower bound for lemma
\ref{sevenPackLemma} and the conjecture.

\topic{Application of Packing Lemma}

So we have this packing lemma, which is fairly straightforward; how can
we use it?

Given a monotone bipartite graph property $\p_f$ on bipartite graphs
$G=(U,W,E)$ with $|U|=|W|=n$, choose $G_1$ to have the
lexicographically smallest degree sequence\defnote{degree sequence}
{The degree sequence of a graph is the list of degrees of the vertices
of the graph, from largest to smallest.} of vertices in $U$, so that
$G_1$ is minimal, no graph in $\p$ has lesser $d^U_{\max}$, and of
those with equal $d^U_{\max}$, none has fewer vertices of this degree.
Similarly, choose $G_2$ lexicographically smallest with
\(\overline{G_2} \not\in \p.\)

We know $G_1$ and $G_2$ don't pack, otherwise 
\(\overline{G_2} \subseteq G_1 \in \p\).

We have several bounds on $D_R(f)$:
\begin{eqnarray}
D_R(f) & \ge & \overline{d}(G_1)n, \label{dBarBound} \\
D_R(f) & \ge & c \frac{d^U_{\max}(G_1)}{\overline{d}(G_1)}n. \label{dMaxBound}
\end{eqnarray}
Bound (\ref{dBarBound}) says that $D_R$ is at least the number of edges in
$G_1$, which is trivial since $G_1$ is minimal.
Bound (\ref{dMaxBound}) is Yao's lemma, shown in the previous lecture.  It also
holds if $G_2$ replaces $G_1$ and/or $W$ replaces $U$, a fact we shall use.

Bound (\ref{dBarBound}) implies that $\overline{d}(G_1) \le D_R/n$, and thus
(\ref{dMaxBound}) implies
\[ d^U_{\max}(G_1) \le \frac{D_R}{cn}\overline{d}(G_1)
\le \frac{1}{c} \left(\frac{D_R}{n}\right)^2,\]
and similarly for $W$ and $G_2$ possible replacing $U$ and $G_1$, 
respectively.  Since $G_1$ and $G_2$ don't pack,
lemma \ref{sevenPackLemma} implies that 
\[\frac{2}{c^2} \left(\frac{D_R}{n}\right)^4
\ge d^U_{\max}(G_1)d^W_{\max}(G_2) + d^U_{\max}(G_2)d^W_{\max}(G_1) 
> n,\]
which in turn implies that
\(D_R \ge \left(\frac{c^2n^5}{2}\right)^{1/4} = \Omega(n^{5/4})\).

Thus any non-trivial bipartite graph property has $D_R = \Omega(n^{5/4})$.

\topic{An Improved Packing Lemma}
To improve this result, we need an improved packing lemma:

\begin{lemma}
Let $G_1$, $G_2$ be bipartite graphs.  If
\begin{eqnarray}
d^U_{\max}(G_1) \overline{d}(G_2) & < & \frac{n}{100}, \label{sevenA}\\
d^U_{\max}(G_2) \overline{d}(G_1)  & < & \frac{n}{100}, \label{sevenB}
\end{eqnarray}
and
\begin{equation}
d^W_{\max}(G_1), d^W_{\max}(G_2) < \frac{n}{1000\log n}, \label{sevenC}
\end{equation}
then $G_1$ and $G_2$ pack.
\end{lemma}

(The condition (\ref{sevenC}) is a technical condition,
needed for the proof but not truly a restriction.  In particular,
if (\ref{sevenC}) is violated, we will see that Yao's lemma
gives an immediate lower bound of $\Omega(n^{3/2}/\sqrt{\log n})$ on $D_R$.)

\begin{proof}
This proof is somewhat more complicated, we sketch the proof.  In
particular, we omit some final computations.

We have $G_1=(U_1,W_1,E_1)$ and $G_2=(U_2,W_2,E_2)$.  We will assume
the above conditions, and show that if we fix a random relabeling $f$
of $W_1$, with non-zero probability there is a relabeling $g$ of
$U_1$ so that $G_1$ relabeled by $f$ and $g$ shares no edges with
$G_2$.

In spirit the idea is initially similar to the previous packing lemma.
There we showed that from the standpoint of a given vertex $u$, if
after ruling out neighbors of neighbors of $u$ there was a vertex
left, we could swap the labels of $u$ and the vertex and possible
reduce the number of edge overlaps.  Thus we showed roughly that the
product of the maximum degree in $U$ and maximum degree in $W$ was at
least $n$.

Here, since the vertices of $W_1$ have been randomly mapped onto the
vertices of $W_2$, the neighbors of $u$ in $G_1$ are mapped onto an
essentially random set of size at most $d_{\max}(G_1)$ in $W_2$.
Since the set in $W_2$ is essentially random, we will be able to show
(using the technical condition) that the size of its neighbor set is
\[c|W_2| \overline{d}(G_2)
 = cd_{\max}(G_1)\overline{d}(G_2) = cn/100\] with probability at most
$\frac{1}{2n}$.  Thus with probability at least $1/2$, for each $u$
there will be less than $n/2$ $u'$ ruled out as possible images under
$g$.  We will also show that the existence of $g$ is equivalent to the
existence of a perfect matching connecting each $u$ with a possible
image $u'$, and thus the existence of $n/2$ possible images of each
$u$ is sufficient to guarantee the existence of $g$.

So fix a relabeling $f$ of $W_1$ uniformly at random.  When will
there be a $g$ relabeling $U_1$ so that no edges are shared?  The
constraint is that if an edge $(u,v)$ is in $E_1$, then the edge
$(g(u), f(v))$ is not in $E_2$.  Thus for each $u\in U_1$ we must find
a $u'\in U_1$ such that $N_{G_2}(u') \cap f(N_{G_1}(u))$ is empty.
($N_G(v)$ denotes the neighboring vertices of $v$ in $G$.)  The only
additional constraint is that each $u$ have a unique such $u'$.

In other words, if we define a bipartite graph \(H = (U_1,U_1,F)\), where
\[F=\{(u,u') : N_{G_2}(u') \cap f(N_{G_1}(u))=\emptyset\},\]
then $g$ exists iff $F$ has a perfect matching.

The Frobenius-Konig-Hall theorem\index{Theorem, Frobenius-Konig-Hall} states
that a bipartite graph $G=(U,W,E)$ has a perfect matching iff for every vertex
set $X\subseteq U$ we have $|N(X)| \ge |X|$.  We don't use this in full
generality, rather we use a consequence.  Namely, if $d_{\min}(G) \ge n/2$,
then $G$ has a perfect matching.  This follows from the FKH theorem as
follows: any set $X$ violating is of size at greater than $n/2$.  But then any
vertex in $W$ has some edge into $X$, since $U-X$ is not big enough to contain
all the edges out of any vertex in $W$.  Thus all vertices in $W$ are
neighbors of $X$.

(Just for fun, note that we can also use the previous lemma to show
this.  Namely if $d_{\min}(G)\ge n/2$, then $\overline{G}$ (with
$d_{\max}(\overline{G})\le n/2$) and a perfect matching
(with $d_{\max}(G)=1$) pack.)

Now $F$ is a random graph, but not in the usual sense.  We will 
argue that with non-zero probability the minimum degree of $F$ is
at least $n/2$, so that it has a perfect matching, and $g$ exists.

To bound the minimum degree of $H$ from below, we ask how many edges
from a vertex $u$ can be excluded.  An edge $(u,u')$ is excluded if
there is a $(u,w) \in E_1$ with $(u',f(w))\in E_2$.  The idea is that
the the number of such $(u,w)$ is bounded by $d^U_{\max}(G_1)$, while
for a given $w$ the number of such $(u',w)$ is around
$\overline{d}(G_2)$, so that for a given $u$, the number of excluded
$u'$ (which must be at least $n$) is at most around the product of
these two.  (By reversing the roles of $G_1$ and $G_2$, we can bound
the number of edges $(u,u')$ into a given $u'$ which are excluded,
thus ensuring $d_H(u')$ is also at least $n/2$ for vertices in the
second part of $H$.)

By the definition of $F$,
\begin{eqnarray}
\#\{u' : (u,u') \not\in H\}
& \le & \left|N_{G_2}\left(f\left(N_{G_1}(u)\right)\right)\right| \nonumber\\
& \le & \sum_{u'\in f\left(N_{G_1}(u)\right)} d_{G_2}(u'). \label{sevenF}
\end{eqnarray}

Thus the probability that $d_H(u) < n/2$ (for $u$ in the first part of
the bipartite graph) is bounded by the probability that (\ref{sevenF})
is greater than $n/2$.  (The case for the second part is similar, and
we omit it.)  The point is that $f\left(N_{G_1}(u)\right)$ is
essentially a random subset of $W_2$ of size at most
$d^U_{\max}(G_1)$, so the sum of the degrees of vertices in $W_2$
should be bounded by $O\left(\overline{d}(G_2) d^U_{\max}\right)$ with
high probability.  In the full proof one shows that for a given $u$
the probability of $D_H(u) \ge n/2$ is at most $\frac{1}{2n}$, so that
the probability of all $u$ having degree less than $n/2$, and thus of
a matching, and the consequent $g$, existing, is at least 1/2.

Here we show exactly what computations we are leaving out:
If we define
\[ \omega_i = \frac{d_{G_2}(w_i)}{n \overline{d}(G_2)}\ \ (w_i\in W_2)\]
then $\sum_i \omega_i = 1$, $\omega_i \ge 0$, and
we want to bound the probability that 
\[ \sum_{i\in S} \omega_i > \frac{1}{2\delta}, \]
where $\delta = \overline{d}(G_2)$ and
$S$ is a set chosen uniformly at random from sets of some
size at most $d^U_{\max}(G_1)$, which is bounded by
$\frac{n}{100\delta}$ by hypothesis.

The average value of the $\omega_i$ is $1/n$, so the expected value of
the sum is $\frac{1}{100\delta}$.  Thus unless the $\omega_i$ are
highly concentrated, which the technical condition $d^W_{\max}(G_1) <
\frac{n}{1000\log n}$ prevents, the condition will hold.
We omit the details of the computation.
\end{proof}

With this improved packing lemma, and the conditions (as before)
\begin{eqnarray}
\frac{D_R}{n} & \ge & \overline{d}(G_1), \label{sevenD}\\
\frac{D_R}{n} & \ge & c \frac{d^U_{\max}(G_1)}{\overline{d}(G_1)}
\label{sevenE}
\end{eqnarray}
(and the corresponding conditions with $W$ and $G_2$ possibly replacing
$U$ and $G_1$, respectively), we can show an improved bound.  (Recall
that the first condition is essentially the trivial lower bound on $D_R$,
while the second is Yao's lemma.)

Specifically, if the technical condition 
\(d^W_{\max}(G_1) < \frac{n}{1000\log n}\)
(or any of the equivalent technical conditions)
of the improved packing lemma
are violated, then by conditions (\ref{sevenD}) and (\ref{sevenE})
\(\frac{D_R^2}{n^2} > \Omega\left(\frac{n}{\log n}\right)\), so
\(D_R = \Omega(n^{3/2}/\sqrt{\log n})\).

Otherwise (as $G_1$ and $G_2$ don't pack) one of the other conditions is
violated.  We assume without loss of generality that
it is (\ref{sevenA}): 
\(d^U_{\max}(G_1) \overline{d}(G_2)  <  \frac{n}{100}\).  Together
with the above two conditions, this gives 
\[\left(\frac{D_R}{n}\right)^3 
\ge c\overline{d}(G_2)d^U_{\max}(G_1)
\ge \frac{cn}{100}.\]

Thus $D_R = \Omega(n^{4/3})$.

It seems possible that this bound could be pushed a bit higher, perhaps
to $n^{3/2}$.  Currently this is the best lower bound known, and the
best upper bounds known are $\Omega(n^2)$.
\lecture{Randomized Complexity of Tree Functions --- Lower Bounds}

For any non-trivial monotone graph property $\p_f$ on graphs with $n$
nodes we have seen that $D(f) = \Omega(n^2)$, $D_R(f) =
\Omega(n^{3/2})$.  

In this lecture we discuss tree functions --- functions with formulas
in which each variable occurs exactly once.  We already know that for
any tree function $f$, $D(f) = n$, and we previously saw a tree
function $f_0$ (represented by a complete binary tree with alternating
and and or-gates) with $D_R(f_0) = O(n^{0.75...})$.

We show a lower bound on $D_R(f)$ for tree functions, which we use to deduce 
that
\begin{itemize}
\item $D_R(f_0) = \Omega(n^{0.75...})$, and
\item $D_R(f) = \Omega(n^{0.51})$ for any tree fn.\ $f$.
\end{itemize}

\topic{Generalized Costs}
The most natural thing to consider for proving a lower bound on the
complexity of a tree function $f(x,y) = g(x)\circ h(y)$ (with $x\in
\{0,1\}^{n-i}$, $y\in \{0,1\}^{i}$, and $\circ\in\{\wedge,\vee\}$)
is a top-down induction.
Unfortunately we don't know how to get this to work.

Instead, Saks and Wigderson have looked at a bottom-up induction, in
which a gate with two immediate inputs is replaced by a single input.
First $f$ is expressed as $f(x,y,w) = f'(x \circ y, w)$ (with $x\in
\{0,1\}$, $y\in \{0,1\}$, $w\in\{0,1\}^{n-2}$, and
$\circ\in\{\wedge,\vee\}$), and then a lower bound on $f$ is given by
a corresponding lower bound on $f'(v,w)$.

For this technique to work, we need to keep track of the fact that
discovering $v$, which represents $x\circ y$, is somehow more
expensive than just querying a bit.  To do this, we generalize our
notion of cost.  We associate two costs $c_0(x_i)$ and $c_1(x_i)$ with
each variable $x_i$ which represents a bit of the input to $f$.  The
cost $c_0$ represents the cost to discover that $x_i$ is 0, while the
cost $c_1$ represents the cost to discover that $x_i$ is 1.  With such
a cost function $c$, we define
\[D_R(f, c) = \min_p \max_x \sum_T p_T \delta(T,x,c),\]
where $\delta(T,x,c) = \sum_{i\in S,x_i=0} c_0(x_i) + \sum_{i\in
S,x_i=1} c_1(x_i)$, with $S$ the set of variables queried by $T$ on
input $x$.

So we are given a function $f$ with a set of costs $c$.  To show a
lower bound on $D_R(f,c)$ we choose the function $f'$ so $f(x,y,w) =
f'(x\circ y,w)$, and we choose a set of costs $c'$ for the inputs of
$f'$ such that we can show $D_R(f,c) \ge D_R(f',c')$, thus inductively
generating a lower bound for $D_R(f,c)$.  We will assume that $\circ =
\wedge$; the case $\circ = \vee$ is symmetric.

So, leaving the choice of $c'$ unspecified as yet, we have $f$, $c$,
$f'$, and $c'$.  We want to show $D_R(f,c) \ge D_R(f',c')$.  The
min-max theorem says $D_R(f, c)$ is also equal to
\[\max_q \min_T \sum_x q_x \delta(T,x,c),\]
that is, we can also obtain $D_R(f,c)$ by choosing the worst input
distribution, and then the best deterministic algorithm for that
distribution.  Thus to show $D_R(f',c')$ is at most $D_R(f,c)$, we
will assume a worst case distribution $q^*$ of inputs to $f'$,
and we show that there exists a $T'$ for $f'$ such that 
$\sum_{x} q^*_x \delta(T', x, c') \le D_R(f, c)$.

So we have the worst case distribution $q^*$ for $f'$, and we want to
show the existence of a $T'$ that does well on $q^*$.  We will map
$q^*$ to a distribution $q$ on the inputs to $f$, so that there exists
an algorithm $T^*$ for $f$ such that $\sum_x q_x\delta(T^*, x, c) \le
D_R(f,c)$.  We know such a $T^*$ exists because $q$ is at worst the
worst-case distribution for $f$, in which case there still exists an
algorithm with expected cost exactly $D_R(f,c)$.  We will then
construct $T'$ based on $T^*$, so that their expected costs on their
respective distributions can be correlated.

The basic idea will be that $T'(v,w)$ will mimic $T^*(x,y,w)$
for some choice of $x$ and $y$ such that $x\wedge y$ = $v$.
When $T^*$ checks a variable in $w$, $T'$ will do the same.
When $T^*$ checks $x$ or $y$, $T'$ may or may not check $v$.

What should the costs $c'$ be?  For variables other than $v$, $c'$
will agree with $c$.  We will wait to determine $c'_0(v)$ and $c'_1(v)$,
choosing them as large as our proof techniques will allow.

What about the distribution $q$?  We define $q(1,1,w) = q^*(1,w)$,
$q(0,1,w) = p_x q^*(0, w)$, and $q(1,0,w) = p_y q^*(0, w)$, where
$p_y$ and $p_x=1-p_y$ will be determined later.

What about $T'$?  Define $T'_y$ on input $(v,w)$ to mimic $T^*$ on
$(1,v,w)$ and $T'_x$ on input $(v,w)$ to mimic $T^*$ on $(v,1,w)$.
Define $T'$ on input $(v,w)$ to run $T'_y$ with probability $p_y$ and
$T'_x$ with probability $p_x$.  That $T'$ is a randomized strategy is
no problem; since $q^*$ is fixed one of the two deterministic strategies
$T'_y$ or $T'_x$ will be at least as good.

We need to find the constraints on $c'_1(v)$ and $c'_0(v)$ which will
allow us to show that the expected cost of $T^*$ is at least that of
$T'$.  For this it suffices to show that the cost of $T^*(v,w)$ for
any $v$ and $w$ is at most $p_y$ times the cost of $T^*(1,v,w)$ plus
$p_x$ times the cost of $T^*(v,1,w)$.

\topic{The Saks-Wigderson Lower Bound}
Now that we have the form of our argument, the rest is essentially a
matter of checking cases.  We should note that although most of the
choices we have made above are straightforward, there is one choice
which in fact anticipates in a clever way what we will need in the
remaining part of the proof.  In particular, we have chosen $T'$ to
run $T'_x$ or $T'_y$ with probability $p_x$ or $p_y$, respectively,
and, not coincidentally, we have chosen the distribution $q$ to map
$(0,w)$ to $(1,0,w)$ or $(0,1,w)$ with probability $p_x$ or $p_y$,
respectively, as well.  This choice bears some consideration.

Returning to our argument, we want to find the conditions under which
the cost of $T'(v,w)$ is at most $p_y$ times the cost of $T^*(1,v,w)$
plus $p_x$ times the cost of $T^*(v,1,w)$.  We consider the various
cases for $T^*$, $v$, $x$, and $y$.

If $v=1$, this reduces to the cost of $T'(1,w)$ being at most the cost
of $T^*(x=1,y=1,w)$.  If $T^*$ queries neither $x$ nor $y$, then this
is clear, since $T'_x$, $T'_y$, and $T^*$ query exactly the same
variables.  Otherwise, if $T^*$ queries only $x$, then $T^*$ pays
$c_1(x)$ while $T'$ pays an expected cost of $p_x c'_1(v)$ (the costs
to query variables in $w$ are again the same); if $T^*$
queries only $y$, then $T^*$ pays $c_1(y)$ while $T'$ pays an expected
cost of $p_y c'_1(v)$; if $T^*$ queries both $x$ and $y$ then $T^*$
pays $c_1(x)+c_1(y)$ while $T'$ pays $c'_1(v)$.  Thus $T^*$ will pay
at least what $T'$ pays provided
\begin{itemize}
\item $p_x c'_1(v) \le c_1(x)$, and
\item $p_y c'_1(v) \le c_1(y)$.
\end{itemize}

The case $v=0$ is a bit more complicated.  In this case, we want the
cost of $T'(0,w)$ to be at most $p_y$ times the cost of $T^*(1,0,w)$
plus $p_x$ times the cost of $T^*(0,1,w)$.  If neither queries $x$ or
$y$ for this $w$ then this is clear.  Otherwise, both query $x$ first
or both query $y$ first.  We consider the case when $x$ is queried
first, the other case being symmetric. 

There are then two cases, depending on whether or not $T^*(1,0,w)$
queries $y$ as well as $x$.  ($T^*(0,1,w)$ will not query $y$, since
$T^*$ is optimal and knows the value of $x\wedge y$ after querying
$x$.)  First assume that only $x$ is queried by $T^*(1,0,w)$.  Then with
probability $p_y$ the cost to $T'$ is the cost to $T^*(1,0,w)$ minus
$c_1(x)$, and with probability $p_x$ the cost to $T'$ is the cost to
$T^*(0,1,w)$ minus $c_0(x)$ plus $c'_0(v)$.  Thus we are fine provided
\begin{itemize}
\item $-p_y c_1(x) - p_x c_0(x) + p_x c'_0(v) \le 0$.
\end{itemize}

Next assume that $x$ and $y$ are queried by $T^*(1,0,w)$.  Then with
probability $p_y$ the cost to $T'$ is the cost to $T^*(1,0,w)$ minus
$c_1(x)$ minus $c_0(y)$ plus $c'_0(v)$, and with probability $p_x$ the
cost to $T'$ is the cost to $T^*(0,1,w)$ minus $c_0(x)$ plus
$c'_0(v)$.  Then we are fine provided
\begin{itemize}
\item $p_y(-c_1(x)-c_0(y)+c'_0(v)) + p_x(-c_0(x)+c'_0(v)) \le 0$.
\end{itemize}

Collecting all of these inequalities, and the symmetric inequalities
for $y$ queried first, we have that $D_R(f,c) \ge D_R(f',c')$ provided
that $c'_0(v)$ and $c'_1(v)$ satisfy the constraints:
\begin{eqnarray*}
c'_1(v) & \le & c_1(x)/p_x, \\
c'_1(v) & \le & c_1(y)/p_y, \\
c'_0(v) & \le & p_y c_1(x)/p_x + c_0(x), \\
c'_0(v) & \le & p_x c_1(y)/p_y + c_0(y), \\
c'_0(v) & \le & p_y(c_1(x)+c_0(y)) + p_x c_0(x),  \\
c'_0(v) & \le & p_x(c_1(y)+c_0(x)) + p_y c_0(y).
\end{eqnarray*}

Choosing $c'_1(v) = c_1(x) + c_1(y)$ forces 
$p_x = \frac{c_1(x)}{c_1(x)+c_1(y)}$ and $p_y = \frac{c_1(y)}{c_1(x)+c_1(y)}$
and yields

\begin{theorem}[Saks-Wigderson\index{Theorem, Saks-Wigderson}]
Let $f$ be a tree function with binary $\wedge$ and $\vee$ gates.
Then $D_R(f) \ge \max\{l^0(f), l^1(f)\}$, where
\begin{eqnarray*}
l^0(x_i) = l^1(x_i) & = & 1, \\
l^1(g\wedge h) & = & l^1(g)+l^1(h), \\
l^0(g\wedge h) & = & \min\{l^0(g)+l^1(h),l^1(g)+l^0(h),
\frac{l^1(g)l^0(g)+l^0(h)l^1(h)+l^1(g)l^1(h)}{l^1(g) + l^1(h)}
\}, \\
l^0(g\vee h) & = & l^0(g)+l^0(h), \\
l^1(g\vee h) & = & \min\{l^1(g)+l^0(h),l^0(g)+l^1(h),
\frac{l^0(g)l^1(g)+l^1(h)l^0(h)+l^0(g)l^0(h)}{l^0(g) + l^0(h)}
\}.
\end{eqnarray*}
\end{theorem}

Applying this to the function $f_0$ (the alternating and/or function
mentioned at the beginning of the lecture) shows that the upper bound
for that function is in fact tight.

Rafi Heiman\index{Heiman} and Avi Wigderson\index{Wigderson} generalize this
theorem to tree functions with arbitrary fan-in gates to show a lower bound of
$\Omega(n^{0.51})$ for the randomized decision tree complexity for an
arbitrary tree function.  See {\it Randomized vs. Deterministic Decision Trees
--- Complexity for Read Once Boolean Functions} by Rafi Heiman and Avi
Wigderson.

\vspace{1in}
(Lecture by Rafi Heiman.)    

\input{techReportIndex.ind}

\end{document}